\documentclass[11pt,a4paper]{article}
\pdfoutput=1

\usepackage{jheppub}

\title{\boldmath Universal Corrections to Non-Extremal Black Hole Entropy in $\mathcal{N} \geq 2$ Supergravity}

\author{Anthony M. Charles and}
\author{Finn Larsen}

\affiliation{Department of Physics and Michigan Center for Theoretical Physics, University of Michigan,\\
450 Church Street, Ann Arbor, MI 48109-1020, USA}

\emailAdd{amchar@umich.edu}
\emailAdd{larsenf@umich.edu}

\abstract{We embed general solutions to 4D Einstein-Maxwell theory into $\mathcal{N} \geq 2$ supergravity and study quadratic fluctuations of the supergravity fields around the background. We compute one-loop quantum corrections for all fields and show that the $c$-anomaly vanishes for complete $\mathcal{N}=2$ multiplets. Logarithmic corrections to the entropy of Kerr-Newman black holes are therefore universal and independent of hole parameters.}

\begin{document}
\maketitle
\flushbottom

\section{Introduction}
\label{sec:introduction}

Logarithmic corrections to the area law for black hole entropy are interesting because they are features of the high energy theory that can be computed systematically in the low energy effective theory~\cite{Banerjee:2010qc,Banerjee:2011jp,Sen:2011ba,Sen:2014aja}. In situations far from the supersymmetric limit there is not yet a microscopic theory of black hole entropy so in this setting logarithmic corrections provide a valuable target for future progress. The most promising arenas for such future developments are nonsupersymmetric black hole solutions to theories with a lot of supersymmetry. The goal of this paper is to compute logarithmic corrections to the entropy of these black holes. 

Logarithmic corrections are derived from quantum determinants over quadratic fluctuations around the black hole background~\cite{Hawking:1976ja}. All fields in the theory fluctuate so the results depend on the theory through its matter content and couplings. Concretely, we consider well-known black holes from general relativity in four dimensions such as the Kerr-Newman solutions to the Einstein-Maxwell theory but we embed these solutions in $\mathcal{N} \geq 2$ SUGRA. The matter content is specified by the host SUGRA which also determines the nonminimal couplings between the matter and the background. 

We find that it is useful to organize the matter in multiplets of $\mathcal{N} = 2$ SUGRA even in the presence of a black hole background that breaks supersymmetry completely and also when $\mathcal{N} > 2$. Indeed, this organization diagonalizes the problem in the sense that different $\mathcal{N} = 2$ multiplets decouple. Furthermore, with our embedding the field equations for quadratic fluctuations of such multiplets depend only on the $\mathcal{N}=2$ field content and not on couplings encoded in the prepotential. 

The one-loop quantum corrections computed in heat kernel regularization are presented as usual as short distance expansions with coefficients that are invariants formed from the curvature \cite{Birrell:1982,Vassilevich:2003xt}. The four derivative terms that we focus on take the form
\begin{equation}
a_4(x) = \frac{c}{16\pi^2}W_{\mu\nu\rho\sigma}W^{\mu\nu\rho\sigma} - \frac{a}{16\pi^2}E_4~,
\end{equation}
where $W_{\mu\nu\rho\sigma}$ is the Weyl tensor and $E_4$ is the 4D Gauss-Bonnet invariant. The values of the coeffiicients $c,a$ are nonstandard because they are for fields with nonminimal couplings specified by $\mathcal{N}\geq 2$ SUGRA. Our results for $c,a$ are somewhat complicated for bosons and fermions separately but we find that the sum gives $c=0$ for any values of $n_V, n_H$, and $\mathcal{N}$. That is very surprising, at least to us. 

The heat kernel coefficient $a_4(x)$ encodes the trace anomaly of the stress tensor which in turn determines the logarithmic correction to the black hole hole entropy in the limit where all parameters with the same length dimension are taken large at the same rate. For BPS black holes there is only one length scale, identified as the scale of the near horizon AdS$_2\times S^2$. In this situation there are no dimensionless ratios so the coefficient of the logarithmic correction is a pure number given by~\cite{Keeler:2014bra}
\begin{equation}
\label{eq:deltas_ads2xs2}
\delta S = \frac{1}{12} \left( 23 - 11 (\mathcal{N}-2) - n_V + n_H\right) \log A_H~.
\end{equation}
For example, BPS black holes in $\mathcal{N}=4$ SUGRA have vanishing logarithmic corrections to their black hole entropy. 

Non-extremal black holes are characterized by dimensionless quantities such as the charge/mass ratio $Q/M$ and the angular momentum quantum number $J/M^2$. For such black holes the coefficient in front of $\log A_H$ is expected to depend on these dimensionless variables. This expectation has proven correct in the case of Kerr-Newman black hole solutions to Einstein-Maxwell theory~\cite{Bhattacharyya:2012wz}. The way this comes about is that fluctuations of the metric and vector fields and additional minimally coupled fields all contribute to the $c$ coefficient, and the curvature invariant $W_{\mu\nu\rho\sigma}W^{\mu\nu\rho\sigma}$ is a complicated function of $Q/M$ and $J/M^2$ after integration over the black hole geometry. 

Our main result is that when the Kerr-Newman black holes are interpreted instead as solutions to $\mathcal{N}\geq 2$ SUGRA the coefficient $c = 0$. In this situation the logarithmic correction is much simpler: (\ref{eq:deltas_ads2xs2}) remains valid for \emph{all} of these black holes (modulo integer corrections due to zero modes and ambiguities in the ensemble). There is no dependence on the parameters that deform the black hole off extremality.

This paper is organized as follows.  In~\S\ref{sec:background} we take solutions to Einstein-Maxwell theory in 4D and embed them into $\mathcal{N}=2$ SUGRA. We determine the quadratic fluctuations around this background and extend these results to general $\mathcal{N} \geq 2$ SUGRA.  In~\S\ref{sec:heatkernel} we review the heat kernel method for determining one-loop quantum corrections to the background fields, and then compute the first three heat kernel coefficients for each $\mathcal{N}=2$ multiplet.  Finally, in~\S\ref{sec:discussion} we tabulate the results for the trace anomaly and show that the logarithmic corrections to black hole entropy are independent of black hole parameters.  As a concrete example we consider the non-extremal Kerr-Newman black hole.

%%%%%%%%%%%%%%%%%%%%%%%%%%%%%%%%%%%%%%%%%%%%%%%%%%%%%%%%%%%%%%

\section{The Background Solution and its Fluctuations}
\label{sec:background}
In this section we embed an arbitrary solution to the $D=4$ Einstein-Maxwell theory into $\mathcal{N} \geq 2$ SUGRA. We then derive the equations of motion for quadratic fluctuations around this background. 

%%%%%%%%%%%%%%%%%%%%%%%%%%%%%%%%%%%%%%%%%%%%%%%%%%%%%%%%%%%%%%

\subsection{The Background Solution: Einstein-Maxwell}
\label{subsec:background_einmax}

The starting point is a solution to the standard $D=4$ Einstein-Maxwell theory
\begin{equation}
\label{eq:aa}
	\mathcal{L} = \frac{1}{2\kappa^2} \left( R  -  \frac{1}{4} F_{\mu\nu} F^{\mu\nu} \right)~,
\end{equation}
where $\kappa^2 = 8\pi G$. The geometry thus satisfies the Einstein equation 
\begin{equation}
\label{eq:ab}
	R_{\mu\nu} - \frac{1}{2} g_{\mu\nu} R = \kappa^2 T_{\mu\nu}~,
\end{equation}
where the energy-momentum tensor is
\begin{equation}
\label{eq:ac}
	T_{\mu\nu} =  - 2 \frac{\delta S}{\delta g^{\mu\nu}} = \frac{1}{2\kappa^2} \left( F_{\mu\rho} F_\nu^{~\rho} - \frac{1}{4}g_{\mu\nu} F_{\rho\sigma} F^{\rho\sigma}\right)~.
\end{equation}
The field strength $F_{\mu\nu}$ satisfies Maxwell's equation and the Bianchi identity which we combine into the complex equation
\begin{equation}
\label{eq:ad}
	\nabla_\mu F^{+\mu\nu} =0~.
\end{equation}
We introduce the self-dual and anti-self-dual parts of the field strength as
\begin{equation}
\label{eq:ae}
	F^{\pm}_{\mu\nu} = \frac{1}{2}(F_{\mu\nu} \pm \widetilde{F}_{\mu\nu})~,
\end{equation}
with
\begin{equation}
\label{eq:af}
	\widetilde{F}_{\mu\nu} = - \frac{i}{2} \varepsilon_{\mu\nu\rho\sigma}F^{\rho\sigma}~.
\end{equation}
We will not specify the solution explicitly at this point but it may be useful to have in mind that we will later consider the Kerr-Newman black hole. 

%%%%%%%%%%%%%%%%%%%%%%%%%%%%%%%%%%%%%%%%%%%%%%%%%%%%%%%%%%%%%%

\subsection{$\mathcal{N}=2$ SUGRA: Notation}
\label{subsec:background_notation}

We want to interpret the background as a solution to $\mathcal{N}=2$ SUGRA with matter in the form of $n_V$ $\mathcal{N}=2$ vector multiplets and $n_H$ $\mathcal{N}=2$ hypermultiplets. The most difficult step will be the bosonic fields in the gravity and vector multiplets so in this section we focus on those. We will return to the hyper multiplets and all the fermions in the next subsection. 

The bosonic action of $\mathcal{N}=2$ SUGRA coupled to $n_V$ vector fields is
\begin{equation}
\label{eq:ba}
\mathcal{L}  = - \frac{1}{2\kappa^2}R - g_{\alpha{\bar \beta}} \nabla^\mu z^{\alpha} \nabla_\mu z^{\bar\beta}
+ \frac{1}{2} {\rm Im} \left[ \mathcal{N}_{IJ} F^{+I}_{\mu\nu}F^{+\mu\nu J}\right]~.
\end{equation}
The index $\alpha=1,\ldots, n_V$ enumerates the complex scalar fields $z^{\alpha}$ in the vector multiplet. The label $I=0,\ldots, n_V$ for the vector fields has an additional value because the $\mathcal{N}=2$ gravity multiplet includes the graviphoton. 

The holomorphic prepotential $F(X)$ is homogeneity two with respect to the projective coordinates $X^I$. It has derivatives denoted $F_{I}=\partial_IF$, $F_{IJ}=\partial_I\partial_J F$, etc., and specifies the coupling between vectors and scalars as
\begin{equation}
\label{eq:bb}
\mathcal{N}_{IJ} = \mu_{IJ} + i\nu_{IJ}  = {\bar F}_{IJ} + i { N_{IK} X^K N_{JL} X^L\over N_{NM} X^N X^M}~,
\end{equation}
where $N_{IJ} = 2{\rm Im} F_{IJ}$. The K{\"a}hler metric is
\begin{equation}
\label{eq:bc}
g_{\alpha{\bar \beta}} = \partial_\alpha \partial_{\bar\beta} \mathcal{K}~,
\end{equation}
where the K{\"a}hler potential is
\begin{equation}
\label{eq:bd}
\mathcal{K} = i (X^I {\bar F}_I - F_I {\bar X}^I )~.
\end{equation}

The K{\"a}hler covariant derivatives 
\begin{equation}\begin{aligned}
\label{eq:be}
\nabla_\alpha X^I &= (\partial_\alpha +  \frac{1}{2} \kappa^2 \partial_\alpha \mathcal{K})X^I~, \\
{\bar\nabla}_{\bar\alpha} X^I &= (\partial_{\bar\alpha} -  \frac{1}{2} \kappa^2 \partial_{\bar\alpha} \mathcal{K})X^I = 0~,
\end{aligned}\end{equation}
relate the true motion in moduli space to the projective parametrization. The K{\"a}hler weights are such that
\begin{equation}
\label{eq:bg}
Z^I (z)= e^{-\frac{1}{2}\kappa^2\mathcal{K}}X^I(z)
\end{equation}
is purely holomorphic $\partial_{\bar\alpha}Z^I=0$. The projective coordinates can be normalized such that
\begin{equation}
\label{eq:bh}
N_{IJ} X^I {\bar X}^J = - i (F_J {\bar X}^J - X^J {\bar F}_J )  = - {1\over\kappa^2}~.
\end{equation}

%%%%%%%%%%%%%%%%%%%%%%%%%%%%%%%%%%%%%%%%%%%%%%%%%%%%%%%%%%%%%%

\subsection{The Background Solution: $\mathcal{N} \geq 2$ SUGRA} 
\label{subsec:background_n=2sugra}

The background solution to Einstein-Maxwell theory specifies the geometry and a single field strength. The corresponding solution to $\mathcal{N}=2$ SUGRA has the same geometry but the matter fields
\begin{equation}\begin{aligned}
\label{eq:ca}
z^\alpha & = {\rm const}~,\cr
F^{+I}_{\mu\nu} & = X^I F^+_{\mu\nu}~. 
\end{aligned}\end{equation}
We need to verify that this in fact is a solution to $\mathcal{N}=2$ SUGRA. 

The scalars are constant so their derivatives do not contribute to the EM tensor:
\begin{equation}
\label{eq:cb}
T_{\mu\nu} = -2 {\rm Im} \left[ {\cal N}_{IJ} (F^{+I}_{\mu\lambda}F^{-\lambda J}_\nu - {1\over 4} g_{\mu\nu} F^{+I}_{\lambda\sigma}F^{-J\lambda\sigma}) \right]  = -2 \nu_{IJ} \left(F^{+I}_{\mu\lambda}F^{-\lambda J}_\nu - {1\over 4} g_{\mu\nu} F^{+I}_{\lambda\sigma}F^{-J\lambda\sigma}\right)~.
\end{equation}
The expression simplifies for the EM tensor of the matter (\ref{eq:ca})  
\begin{equation}
\label{eq:cc}
- 2\nu_{IJ} X^I {\bar X}^J = i {\cal N}_{IJ} X^I {\bar X}^J  + {\rm c.c} = i F_{IJ} X^I{\bar X}^J+ {\rm c.c} 
= i (F_J {\bar X}^J - X^J {\bar F}_J ) = {1\over\kappa^2}~,
\end{equation}
due to (\ref{eq:bb}) for $\mathcal{N}_{IJ}$, homogeneity $2$ of the prepotential, and then the normalization condition (\ref{eq:bh}). 
Our result therefore becomes
\begin{equation}\begin{aligned}
\label{eq:cd}
T_{\mu\nu} & = {1\over\kappa^2} \left( F^{+}_{\mu\lambda} F_\nu^{-\lambda} - {1\over 4} g_{\mu\nu} F^{+}_{\lambda\sigma} F^{-\lambda\sigma}\right) = {1\over 2\kappa^2} \left( F_{\mu\lambda} F_\nu^{~\lambda} - {1\over 4} g_{\mu\nu} F_{\lambda\sigma} F^{\lambda\sigma}\right)~. 
\end{aligned}\end{equation}
This is the same as (\ref{eq:ac}) for the Einstein-Maxwell theory so the Einstein equation for $\mathcal{N}=2$ SUGRA is satisfied with unchanged geometry. 

The combined Maxwell-Bianchi equation in $\mathcal{N}=2$ SUGRA
\begin{equation}
\label{eq:ce}
\nabla_\mu \left( {\cal N}_{IJ} F^{+I\mu\nu} \right) =0~
\end{equation}
is automatically satisfied because the background satisfies the Maxwell-Bianchi equation (\ref{eq:ad}). The dependence of $\mathcal{N}_{IJ}$ and $F^{I\mu\nu}$ on the scalar fields introduces no spacetime dependence since the scalars are constant. 

The scalar field equations are not automatic even though the scalars are constant because the vector fields act as a source unless
\begin{equation}\begin{aligned}
\label{eq:cf}
{\partial {\cal N}_{IJ}\over\partial z^\alpha}  F^{+I}_{\mu\nu}F^{+\mu\nu J} & ={\partial {\cal N}_{IJ}\over\partial  z^{\bar\alpha}}  F^{+I}_{\mu\nu}F^{+\mu\nu J} = 0 ~.
\end{aligned}\end{equation}
The anti-holomorphic condition 
\begin{equation}\begin{aligned}
\label{eq:cg}
{\partial {\cal N}_{IJ}\over\partial z^{\bar\alpha}}  Z^I Z^J  = 0
\end{aligned}\end{equation}
is obvious: move the holomorphic coordinates $Z^I$ under the derivative and use $\mathcal{N}_{IJ} Z^I Z^J = F_I Z^I = 2F$ to find an antiholomorphic derivative that vanishes because it acts on the holomorphic prepotential. The holomorphic condition is almost as simple:
\begin{equation}
\label{eq:ch}
{\partial {\cal N}_{IJ}\over\partial z^\alpha}  Z^I Z^J = \partial_\alpha F_I Z^I - F_I \partial_\alpha Z^I  
=    (F_{IJ} Z^I - F_J )\partial_\alpha Z^J   = 0 ~.
\end{equation}
We used $\mathcal{N}_{IJ}Z^J=F_I$ again and then $F_{IJ} Z^I =F_J$ from homogeneity of the prepotential. 

At this point we have completed the verification that a solution to Einstein-Maxwell remains a solution when embedded in $\mathcal{N}=2$ SUGRA through (\ref{eq:ca}).  Any additional fields in $\mathcal{N} > 2$ SUGRA must all appear quadratically.  The further embedding from $\mathcal{N}>2$ into $\mathcal{N} = 2$ SUGRA is therefore automatic.

%%%%%%%%%%%%%%%%%%%%%%%%%%%%%%%%%%%%%%%%%%%%%%%%%%%%%%%%%%%%%%

\subsection{Quadratic Fluctuations: the Action}
\label{subsec:background_quadfluct_action}
The quantum corrections to the black hole entropy are determined by the spectrum of quadratic fluctuations around the background. 

We first consider the general matter equations of motion derived from the action (\ref{eq:ba})
\begin{equation}\begin{aligned}
\label{eq:da}
\nabla^\mu \left( g_{\alpha{\bar\beta}} \nabla_\mu z^\alpha \right)- {i\over 4} {\partial{\cal N}_{IJ}\over\partial{\bar z}^{\bar\beta}} F_{\mu\nu}^{+I}F^{+J\mu\nu}  
+ {i\over 4} {\partial\overline{\cal N}_{IJ}\over\partial{\bar z}^{\bar\beta}} F_{\mu\nu}^{-I}F^{-J\mu\nu} &= 0~,  \cr
i \nabla_\mu \left( {\cal N}_{IJ}F^{+J\mu\nu} - \overline{\cal N}_{IJ}F^{-J\mu\nu}\right) & = 0~, \cr
\nabla_\mu \left( F^{+J\mu\nu} - F^{-J\mu\nu}\right) & = 0~. 
\end{aligned}\end{equation}
The last two lines are the Maxwell-Bianchi equations. Linearizing around the background these equations become
\begin{equation}\begin{aligned}
\label{eq:db}
{\cal N}_{IJ} \nabla_\mu \delta F^{+J\mu\nu} + (\nabla_\mu \delta {\cal N}_{IJ} ) F^{+J\mu\nu}  - 
\overline{\cal N}_{IJ} \nabla_\mu\delta F^{-J\mu\nu} - 
(\nabla_\mu\delta \overline{\cal N}_{IJ}) F^{-J\mu\nu}  & = 0~, \cr
\nabla_\mu \left( \delta F^{+J\mu\nu} - \delta F^{-J\mu\nu}\right) & = 0 ~,
\end{aligned}\end{equation}
where unvaried fields (without $\delta$) are evaluated on the background. The background fields satisfy (\ref{eq:ca}) so the variation simplifies 
\begin{equation}\begin{aligned}
\label{eq:dc}
(\nabla_\mu \delta {\cal N}_{IJ}) F^{+J\mu\nu} &= (\nabla_\alpha {\cal N}_{IJ}) X^J \nabla_\mu \delta z^\alpha F^{+\mu\nu}+
\nabla_{\bar\alpha} {\cal N}_{IJ} X^J \nabla_\mu \delta z^{\bar\alpha} F^{+\mu\nu}\cr
& = (\nabla_\alpha F_I - {\cal N}_{IJ} \nabla_\alpha X^J) \nabla_\mu \delta z^\alpha F^{+\mu\nu} 
= -2i\nu_{IJ} \nabla_\alpha X^J \nabla_\mu \delta z^\alpha F^{+\mu\nu}  F^{+\mu\nu} ~.
\end{aligned}\end{equation}
We used symplectic invariance in the form $\nabla_\alpha F_I = \overline{N}_{IJ} \nabla_\alpha X^J$. 

Inserting into (\ref{eq:db}) and simplifying we find
\begin{equation}\begin{aligned}
\label{eq:dd}
\nabla_\mu \left( \delta F^{+I\mu\nu} - \nabla_{\alpha} X^I \delta z^\alpha  F^{+\mu\nu} - \nabla_{\bar\alpha} {\bar X}^I \delta z^{\bar\alpha}   F^{-\mu\nu}\right)=0~.
\end{aligned}\end{equation}
This is a complex equation with imaginary part reducing to the Bianchi identity. 

After linearizing the scalar equation of motion in (\ref{eq:da}) around the background the middle term vanishes due to holomorphicity. The (complex conjugate of) the last term simplifies as
\begin{equation}\begin{aligned}
\label{eq:de}
\delta \left( \nabla_\beta {\cal N}_{IJ} F^{+I}_{\mu\nu} F^{+J\mu\nu} \right)  & = \nabla_{\alpha}\nabla_\beta {\cal N}_{IJ} \delta z^\alpha X^I X^J F^{+}_{\mu\nu} F^{+\mu\nu} + 2\nabla_\beta{\cal N}_{IJ}X^I F^{+}_{\mu\nu} \delta F^{+J\mu\nu}\cr
& = 2\nabla_\beta{\cal N}_{IJ}X^I F^{+}_{\mu\nu} (\delta F^{+J\mu\nu} - \nabla_\alpha X^J \delta z^\alpha F^{+\mu\nu})\cr
& = -4i\nu_{IJ} \nabla_\beta X^I F^{+}_{\mu\nu} (\delta F^{+J\mu\nu} - \nabla_\alpha X^J \delta z^\alpha F^{+\mu\nu})~.
\end{aligned}\end{equation}
We then collect terms and write the linearized scalar equation as
\begin{equation}
\label{eq:df}
g_{\alpha{\bar\beta}} \nabla^2 \delta z^\alpha - \nu_{IJ} \bar{\nabla}_{\bar\beta} {\bar X}^I F^{-}_{\mu\nu} (\delta F^{-J\mu\nu} - {\bar\nabla}_{\bar\alpha} {\bar X}^J \delta z^{\bar\alpha}F^{-\mu\nu}) = 0 ~.
\end{equation}

The linearized equations of motion for the vectors (\ref{eq:dd}) and the scalars (\ref{eq:df}) can both be derived from the single action 
\begin{equation}
\label{eq:dg}
{\cal L} = - g_{\alpha{\bar\beta}} \nabla_\mu \delta z^\alpha  \nabla^\mu \delta z^{\bar\beta} + {1\over 2} \nu_{IJ} ( \delta F^{+I}_{\mu\nu} - \delta X^I F^+_{\mu\nu} )( \delta F^{+J\mu\nu} - \delta X^J F^{+\mu\nu} )+ {\rm c.c}~,
\end{equation}
with $\delta X^I = \nabla_\alpha X^I \delta z^\alpha$. This is a consistency check on the manipulations. 

%%%%%%%%%%%%%%%%%%%%%%%%%%%%%%%%%%%%%%%%%%%%%%%%%%%%%%%%%%%%%%

\subsection{Quadratic Fluctuations: Decoupling}
\label{subsec:background_quadfluct_decoupling}
The action (\ref{eq:dg}) for quadratic fluctuations is concise but the dependence on the K{\"a}hler metric $g_{\alpha{\bar\beta}}$ and the symplectic metric $\nu_{IJ}$ introduces elaborate couplings between the $n_V$ complex scalars $z^\alpha$ and the $n_V+1$ field strengths $F^I_{\mu\nu}$. We can simplify by expanding as 
\begin{equation}\begin{aligned}
\label{eq:dh}
F^{+I}_{\mu\nu} & = X^I F^+_{\mu\nu} + {\bar\nabla}_{\bar\alpha} \bar{X}^I f^{+{\bar\alpha}}_{\mu\nu}\cr
& = \begin{pmatrix} X^I & {\bar\nabla}_{\bar\alpha} \bar{X}^I\end{pmatrix} \begin{pmatrix} F^+_{\mu\nu} \cr f^{+{\bar\alpha}}_{\mu\nu}\end{pmatrix}~.
\end{aligned}\end{equation}
This represents the $n_V+1$ fields $F^I_{\mu\nu}$ as a single graviphoton field $F_{\mu\nu}$ and $n_V$ vector fields $f^\alpha_{\mu\nu}$. 
The complete basis $\{ X^I, \overline{\nabla}_{\bar\alpha} \overline{X}^I \}$ is orthogonal with respect to the metric $\nu_{IJ}$ in the sense that
\begin{equation}
\label{eq:di}
\begin{pmatrix} {\bar X}^I \cr \nabla_\alpha X^I\end{pmatrix} \nu_{IJ} \begin{pmatrix} X^I & {\bar\nabla}_{\bar\beta} \bar{X}^I\end{pmatrix}  = - {1\over 2} \begin{pmatrix} \kappa^{-2} & 0 \cr 0 & g_{\alpha{\bar\beta}}\end{pmatrix}~.
\end{equation}
The component form of the field variations are
\begin{equation}
\label{eq:dj}
\delta F^{+I\mu\nu} = X^I\delta F^{+\mu\nu} + \nabla_\alpha X^I \delta z^\alpha F^{+\mu\nu} + {\bar\nabla}_{\bar\alpha} {\bar X}^I 
f^{+{\bar\alpha}\mu\nu} ~.
\end{equation}
For variations of this form the linearized matter equations (\ref{eq:dd}) and (\ref{eq:df}) become
\begin{equation}\begin{aligned}
\label{eq:dl}
X^I \nabla_\mu \delta F^{+\mu\nu}  + {\bar\nabla}_{\bar\alpha} {\bar X}^I\nabla_\mu \left( f^{+{\bar\alpha}\mu\nu}  - \delta z^{\bar\alpha} F^{-\mu\nu} \right) &= 0~, \cr
\nabla^2 \delta z^\alpha + {1\over 2} F^{-}_{\mu\nu} f^{-\alpha\mu\nu}  &= 0 ~.
\end{aligned}\end{equation}
Orthogonality forces the two terms in the first equation to vanish separately. Thus $\delta F^{+\mu\nu} $ satisfies the standard Maxwell-Bianchi equations (\ref{eq:ad}) also in the presence of a fluctuating scalar. We rename $\delta z^\alpha\to z^\alpha$ and write the remaining equations as
\begin{equation}\begin{aligned}
\label{eq:dm}
\partial_\mu \left(  f^{-\alpha\mu\nu}  -  z^\alpha F^{+\mu\nu} \right) &= 0~, \cr
\nabla^2 z^\alpha + {1\over 2} F^{-}_{\mu\nu} f^{-\alpha\mu\nu}  &= 0 ~.
\end{aligned}\end{equation}
These matter equations are fully decoupled: there is no interaction between the matter multiplet and the SUGRA multiplet (gravity and graviphoton). Also, the $n_V$ vector multiplets do not couple to each other so they can be analyzed independently. We will henceforth suppress the index $\alpha$.

The equations of motion (\ref{eq:dm}) are actually misleading as they stand because, according to the first equation, the antisymmetric vector field $f^{\mu\nu}$ does not satisfy the Bianchi identity: the imaginary part of $f^{-\mu\nu}$ has a source. We can remedy this by the field redefinition 
\begin{equation}
\label{eq:dn}
f^{-\mu\nu}\to -2if^{-\mu\nu} + \bar{z}F^{-\mu\nu}~.
\end{equation}
The transformed field strength $f^{\mu\nu}$ satisfies the Bianchi identity. The general equations (\ref{eq:dm}) become 
\begin{equation}\begin{aligned}
\label{eq:do}
\partial_\mu \left(  f^{\mu\nu}  -   i z F^{+\mu\nu} + i {\bar z}F^{-\mu\nu} \right) &= 0~, \cr
\nabla^2 z - i F^{-}_{\mu\nu} f^{-\mu\nu} + {1\over 2}\bar{z} F^-_{\mu\nu}F^{-\mu\nu}  &= 0~.
\end{aligned}\end{equation}
This is our final result for quadratic fluctuations of a ${\cal N}=2$ vector multiplet around a solution to the Einstein-Maxwell theory. 

We have not yet analyzed the Einstein equation. Linearizing the EM tensor (\ref{eq:cb}) of the ${\cal N}=2$ theory around the background we find
\begin{equation}\begin{aligned}
\label{eq:dq}
\delta \left( \nu_{IJ} F^{+I}_{ac} F^{-Jc}_b \right) &= \delta \nu_{IJ} F^{+I}_{ac} F^{-Jc}_b +   \nu_{IJ} \left(  \delta F^{+I}_{ac} F^{-Jc}_b + {\rm c.c.} \right) \cr
& = - {i\over 2} \nabla_\alpha \left(  {\cal N}_{IJ}  - {\bar{\cal N}}_{IJ}\right) \delta z^\alpha X^I {\bar X}^J F^{+}_{ac} F^{-c}_b + 
\nu_{IJ} \delta F^{+I}_{ac} {\bar X}^J F^{-c}_b + {\rm c.c.} \cr 
& = \nu_{IJ}  {\bar X}^J  ( \delta F^{+I}_{ac} - \nabla_\alpha X^I  \delta z^\alpha F^{+}_{ac} ) F^{-c}_b  + {\rm c.c.} \cr
& = - {1\over 2\kappa^2} \delta (F^{+}_{ac} F^{-c}_b)~, 
\end{aligned}\end{equation}
for variations of the form (\ref{eq:dj}). Thus fluctuations in the geometry are sourced exclusively by the graviphoton or, equivalently, the Einstein equation respects the decoupling of the ${\cal N}=2$ SUGRA multiplet from the vector multiplets. 

%%%%%%%%%%%%%%%%%%%%%%%%%%%%%%%%%%%%%%%%%%%%%%%%%%%%%%%%%%%%%%

\subsection{Quadratic Fluctuations: Completing the Multiplets}
\label{subsec:background_quadfluct_completing}
The full ${\cal N}=2$ supergravity theory generally includes many fields that vanish in the background. The actions of such fields can be computed at quadratic order by taking all other fields to have their background value. This process introduces nonminimal couplings because of the background graviphoton. In the following we examine the various ${\cal N}=2$ multiplets one by one. 

The ${\cal N}=2$ SUGRA multiplet contains the graviton, two gravitini, and the graviphoton. The bosons are governed by the Einstein-Maxwell action (\ref{eq:aa}). The gravitino action is\footnote{We follow the conventions of~\cite{Freedman:2012zz}. The normalization of the Maxwell field strength in (\ref{eq:aa}) is conventional in the gravity literature. The relation between the two conventions for the graviphoton 
is $F^{\text{(FvP)}}_{\mu\nu} = 2F^{\text{(here)}}_{\mu\nu}$.}
\begin{equation}
\label{eq:ea}
{\cal L}_{\rm gravitino} =  - {1\over\kappa^2}\bar{\psi}_{i\mu}\gamma^{\mu\nu\rho}D_\nu\psi_\rho^i  +  \nu_{IJ}\left( 
F^{-I}_{\mu\nu}Q^{-J\mu\nu} + {\rm h.c.} \right) ~,
\end{equation}
where $i=1,2$ enumerates the two gravitini. The Pauli term depends on $Q^{-J\mu\nu} = X^J \bar{\psi}^\mu_i \psi^\nu_j \varepsilon^{ij} + \ldots$. In the background (\ref{eq:ca}) the normalization condition (\ref{eq:di}) then gives 
\begin{equation}
\label{eq:eb}
{\cal L}_{\rm gravitino} =  - {1\over\kappa^2} \bar{\psi}_{i\mu}\gamma^{\mu\nu\rho}D_\nu\psi_\rho^i - {1\over 2\kappa^2} \left( F^-_{\mu\nu}\bar{\psi}^\mu_i \psi^\nu_j \varepsilon^{ij} + {\rm h.c.} \right) ~.
\end{equation}
The sum of this action and the Einstein-Maxwell action (\ref{eq:aa}) is invariant under the ${\cal N}=2$ supersymmetry
\begin{equation}
\label{eq:ec}
\delta\psi^i_\mu = D_\mu \epsilon^i - {1\over 4}{\hat F} \varepsilon^{ij} \gamma^\mu \epsilon_j~,
\end{equation}
where ${\hat F}  = {1\over 2} \gamma_{\mu\nu} F^{\mu\nu}$. 

The ${\cal N}=2$ vector multiplet has one vector field, two gauginos, and one complex scalar. The bosons satisfy the equations of motion (\ref{eq:do}), as we have shown in detail. The gauginos are subject to Pauli terms that couple them to each other and to the gravitinos. However, these couplings appear in the combination
$Q^{-J\mu\nu} = \overline{\nabla}_{\bar{\alpha}}\overline{X}^J \left( \overline{\chi}^{\bar{\alpha}i} \gamma^\mu \psi^{\nu j}\varepsilon_{ij}+ \ldots\right)$, and such terms vanish when contracted with a field strength $F^{-I}_{\mu\nu}$ of the background form (\ref{eq:ca}) because of orthogonality (\ref{eq:di}). Therefore the gauginos are minimally coupled fermions. 

The ${\cal N}=2$ hyper multiplet has two Majorana hyper fermions and four real scalars. The scalars are minimally coupled to gravity. The hyper fermion is acted on by a Pauli term where the metric $\nu_{IJ}$ contracts $F^{-I}_{\mu\nu}$ and $Q^{-J\mu\nu} =X^J \left( {1\over 2}\kappa^2 \overline{\zeta}^A \gamma^{\mu\nu}\zeta^BC_{AB}+ \ldots\right)$. The $Q^{-J\mu\nu}$ is proportional to $X^J$ as it was for the gravitino and again orthogonality leads to a simple result for the quadratic action
\begin{equation}
\label{eq:ed}
{\cal L}_{\rm hyper} = - 2\overline{\zeta}_A  \gamma^\mu D_\mu \zeta^A  - {1\over 2} \left( \overline{\zeta}^A {\hat F} \zeta^B \epsilon_{AB} + {\rm h.c.} \right)~.
\end{equation}
It is sufficient for our purposes to consider each hyper multiplet independently. For a single hyper multiplet the indices $A,B=1,2$ and we can take $C_{AB}=\epsilon_{AB}$. 

%%%%%%%%%%%%%%%%%%%%%%%%%%%%%%%%%%%%%%%%%%%%%%%%%%%%%%%%%%%%%%

\subsection{${\cal N}>2$ SUGRA}
\label{subsec:background_n>2sugra}
We can extend our results for ${\cal N}=2$ SUGRA to theories with ${\cal N}>2$. 

We first embed the background solution to the Maxwell-Einstein theory into SUGRA with ${\cal N}>2$. We pick one of the ${1\over 2}{\cal N}({\cal N}-1)$ gauge fields in the gravity multiplet and designate it as the graviphoton of an ${\cal N}=2$ theory that is identified with the gauge field of the Maxwell-Einstein theory. This embedding defines the background defined earlier as a solution also to ${\cal N}>2$ SUGRA. 

We next organize all fluctuating fields in ${\cal N}=2$ multiplets. The ${\cal N}>2$ symmetry constrains the ${\cal N}=2$ matter content. For example: 

\begin{itemize}
\item A ${\cal N}=4$ theory has $n_V=n_H+1$ because one ${\cal N}=2$ vector is part of the ${\cal N}=4$ supergravity multiplet while each ${\cal N}=4$ matter multiplet is composed of one ${\cal N}=2$ vector and one ${\cal N}=2$ hyper. 

\item The ${\cal N}=6$ theory: $n_V=7$ and $n_H=4$.

\item The ${\cal N}=8$ theory: $n_V=15$ and $n_H=10$.
\end{itemize}

The classification under ${\cal N}=2$ takes all matter fields into account except for the ${\cal N}-2$ gravitini and their superpartners. We refer to these as massive gravitini. A massive gravitino multiplet has one gravitino, two vectors, and one gaugino. The two vectors in the massive gravitino multiplet are minimally coupled vector fields. The background graviphoton field couples the remaining fermions:
\begin{equation}
\label{eq:ee}
{\cal L}_{\rm gravitino} 
= -  {1\over\kappa^2}\bar{\Psi}_{\mu} \gamma^{\mu\nu\rho}D_\nu\Psi_\rho - {2\over\kappa^2}\overline{\lambda}  \gamma^\mu D_\mu \lambda
- {1\over 2\kappa^2} (\bar{\Psi}_{\mu}{\hat F} \gamma^\mu\lambda + {\rm h.c.} ) ~.
\end{equation}
We found this action by reduction of ${\cal N}=8$ SUGRA~\cite{Corley:1999uz} but other approaches give the same result. As a check we verified that in AdS$_2\times S^2$ the fermion fields acquire precisely the conformal weights demanded by the superconformal symmetry of the action.

%%%%%%%%%%%%%%%%%%%%%%%%%%%%%%%%%%%%%%%%%%%%%%%%%%%%%%%%%%%%%%

\section{Heat Kernel Expansion}
\label{sec:heatkernel}

This section reviews the heat kernel approach to the computation of functional determinants of the quadratic field fluctuations, including a few elementary examples.  We then employ this method to calculate heat kernel coefficients for quadratic fluctations around generic solutions to Einstein-Maxwell theory.  We consider each $\mathcal{N}=2$ multiplet in turn.

%%%%%%%%%%%%%%%%%%%%%%%%%%%%%%%%%%%%%%%%%%%%%%%%%%%%%%%%%%%%%%

\subsection{One-Loop Quantum Corrections}
\label{subsec:heatkernel_quantumcorrections}

The action for quadratic fluctuations around a background solution takes the schematic form
\begin{equation}
	\mathcal{S} = -\int d^4 x\,\sqrt{-g}\,\phi_n \Lambda^n_m \phi^m~,
\end{equation}
for some differential operator $\Lambda$ that depends on the background fields such as the metric $g_{\mu\nu}$ and the field strength $F_{\mu\nu}$.  We define the effective action $W$ as
\begin{equation}
	W = \frac{1}{2}\log\det\Lambda~,
\end{equation}
since we can schematically write
\begin{equation}
\label{eq:pathint}
	e^{-W} = \int \mathcal{D}\phi\,e^{-\phi\Lambda\phi} = \frac{1}{\sqrt{\det\Lambda}}~.
\end{equation}
The one-loop quantum corrections we want to compute are encoded in this Euclidean path integral.

The heat kernel of $\Lambda$ is defined by
\begin{equation}
\label{eq:d(s)}
	D(s) = \text{Tr }e^{-s\Lambda} = \sum_i e^{-s\lambda_i}~,
\end{equation}
where $\{\lambda_i\}$ are the eigenvalues of $\Lambda$; we have denoted them as discrete, but they may in practice be continuous.  After UV regularization we can express $W$ in terms of $D(s)$ as
\begin{equation}
\label{eq:w-d(s)}
	W = \frac{1}{2}\sum_i \log\lambda_i = -\frac{1}{2}\int^\infty_{\epsilon^2}ds\,\frac{D(s)}{s}~.
\end{equation}
$D(s)$ is referred to as the ``heat kernel" because we can express it as
\begin{equation}
\label{eq:d(s)-k}
	D(s) = \int d^4x\,\sqrt{-g}\,K(x,x;s)~,
\end{equation}
where the Green's function $K(x,x';s)$ satisfies the heat equation
\begin{equation}
	(\partial_s + \Lambda_x)K(x,x';s)=0~,
\end{equation}
with the boundary condition $K(x,x';0) = \delta(x-x')$.  Inserting (\ref{eq:d(s)-k}) back into (\ref{eq:w-d(s)}), $W$ becomes
\begin{equation}
	W = -\frac{1}{2}\int^\infty_{\epsilon^2}\frac{ds}{s}\int d^Dx\,\sqrt{-g}\, K(x,x;s)~.
\end{equation}
Therefore, in order to compute the one-loop quantum corrections to the theory, we must find the Green's function $K(x,x;s)$ corresponding to the quadratic field operator $\Lambda$.

%%%%%%%%%%%%%%%%%%%%%%%%%%%%%%%%%%%%%%%%%%%%%%%%%%%%%%%%%%%%%%

\subsection{General Method}
\label{subsec:heatkernel_generalmethod}

The Green's function $K(x,x;s)$ corresponding to $\Lambda$ permits a perturbative expansion for small $s$
\begin{equation}
\label{eq:kseries}
	K(x,x;s) = \sum_{n=0}^\infty s^{n-2}a_{2n}(x)~,
\end{equation}
where $\{a_{2n}(x)\}$ are the Seeley-DeWitt coefficients~\cite{DeWitt:102655}.  We want to express these coefficients in terms of the background fields and their covariant derivatives.  To do so, we follow the procedure reviewed in~\cite{Vassilevich:2003xt} and assume that the differential operator $\Lambda$ acting on our quadratic field fluctuations takes the form
\begin{equation}
\label{eq:lambda_def}
	\Lambda^n_m = -(\square)I^n_m - 2(\omega^\mu D_\mu)^n_m - P^n_m~,
\end{equation}
where $\omega^\mu$ and $P$ are matrices constructed from the background fields $\{\phi_n\}$ and $I$ is the identity operator on the space of fields. We denote $\square = D^\mu D_\mu$ where $D_\mu$ indicates the ordinary covariant derivative.  All indices are raised and lowered with the background metric $g_{\mu\nu}$.  The formulae below require that $\Lambda$ be a Hermitian operator, and so the action must be adjusted (up to a total derivative) to make this so.

We can complete the square and rewrite $\Lambda$ as
\begin{equation}
	\Lambda^n_m = - (\mathcal{D}^\mu \mathcal{D}_\mu)I^n_m - E^n_m~,
\end{equation}
where
\begin{equation}
	E = P - \omega^\mu\omega_\mu - \left(D^\mu \omega_\mu\right)~,\quad \mathcal{D}_\mu = D_\mu + \omega_\mu~,
\end{equation}
and we have suppressed the field indices for notational simplicity.  Note that the parentheses in the term $\left(D^\mu \omega_\mu\right)$ indicate that the covariant derivative $D^\mu$ acts only on $\omega_\mu$.  The background connection was assumed to be torsion-free, but this new effective covariant derivative $\mathcal{D}_\mu$ need not be.  The curvature associated with $\mathcal{D}_\mu$ is denoted as
\begin{equation}
\label{eq:omeg}
	\Omega_{\mu\nu} \equiv [\mathcal{D}_\mu,\mathcal{D}_\nu]~.
\end{equation}
With these definitions, the expressions for the first three Seeley-DeWitt coefficients are
\begin{equation}\begin{aligned}
\label{eq:a2n}
	(4\pi)^2a_0(x) &= \text{Tr }I~, \\
	(4\pi)^2a_2(x) &= \text{Tr }E~, \\
	(4\pi)^2a_4(x) &= \text{Tr }\left[\frac{1}{2}E^2 + \frac{1}{12}\Omega_{\mu\nu}\Omega^{\mu\nu} + \frac{1}{180}(R_{\mu\nu\rho\sigma}R^{\mu\nu\rho\sigma} - R_{\mu\nu}R^{\mu\nu})I\right]~. \\
\end{aligned}\end{equation}
We have set $R=0$ in (\ref{eq:a2n}) since the Einstein equation (\ref{eq:ab}) defines a Ricci-flat space.  We have also ignored all total derivatives as they will integrate to zero.  We can in principle go to higher orders and compute $a_{6}(x)$, $a_{8}(x)$, \ldots, but the logarithmic divergences in 4D are determined by $a_4(x)$ and so it is sufficient to compute the heat kernels up to this order.

%%%%%%%%%%%%%%%%%%%%%%%%%%%%%%%%%%%%%%%%%%%%%%%%%%%%%%%%%%%%%%

\subsection{Elementary Examples}
\label{subsec:heatkernel_examples}

To see how these heat kernel coefficient calculations work in practice, we will do the explicit calculations for a few elementary examples.  The methods used naturally generalize for the more complicated interactions analyzed in the following subsection.

%%%%%%%%%%%%%%%%%%%%%%%%%%%%%%%%%%%%%%%%%%%%%%%%%%%%%%%%%%%%%%

\subsubsection{Free Scalar Field}
\label{subsubsec:heatkernel_examples_scalar}

Consider a minimally-coupled scalar field with a mass $m$
\begin{equation}
	\mathcal{L} = -\frac{1}{2}(\partial^\mu \phi)(\partial_\mu \phi) -\frac{1}{2} m^2 \phi^2~.
\end{equation}
The ordinary derivative $\partial_\mu$ is the same as the covariant derivative $D_\mu$ when acting on scalar fields so we can integrate by parts freely.  In the language of~\S\ref{subsec:heatkernel_generalmethod} the differential operator is
\begin{equation}
\label{eq:lambda_scalar}
	\Lambda = -\square + m^2~,
\end{equation}
where $\Lambda$ acts on the scalar field $\phi$.  There are no terms in (\ref{eq:lambda_scalar}) linear in derivatives, so the matrices $I$, $\omega_\mu$, and $E$ are 
\begin{equation}
	I = 1~, \quad \omega_\mu = 0~, \quad E = -m^2~.
\end{equation}
The commutator of two covariant derivatives vanish when acting on scalar fields, and so the curvature is 0:
\begin{equation}
	\Omega_{\mu\nu} = 0~.
\end{equation}
Inserting this data into (\ref{eq:a2n}) we find the Seeley-DeWitt coefficients for the massive scalar field
\begin{equation}\begin{aligned}
\label{eq:a2n_scalar}
(4\pi)^{2} a_0(x) &= 1~, \\
(4\pi)^{2}a_2(x) &= -m^2 ~, \\
(4\pi)^{2}a_4(x) &= \frac{1}{2}m^4 +\frac{1}{180}\left(R_{\mu\nu\rho\sigma}R^{\mu\nu\rho\sigma}-R_{\mu\nu}R^{\mu\nu}\right)~.
\end{aligned}
\end{equation}

%%%%%%%%%%%%%%%%%%%%%%%%%%%%%%%%%%%%%%%%%%%%%%%%%%%%%%%%%%%%%%

\subsubsection{Free Spinor Field}
\label{subsubsec:heatkernel_examples_spinor}

The action for a minimally-coupled Dirac spinor $\psi$ with a (real) mass $m$ in a 4D spacetime is
\begin{equation}
\label{eq:l_h}
	\mathcal{S} = \int d^4x\,\sqrt{-g}\,\bar{\psi}\left( \gamma^\mu D_\mu + m\right)\psi~.
\end{equation}
The gamma matrices $\gamma_\mu$ satisfy the standard commutation relation
\begin{equation}
	\{\gamma_\mu,\gamma_\nu\} = 2 g_{\mu\nu}\mathbb{I}_4~,
\end{equation}
where $\mathbb{I}_4$ is the identity matrix in our Clifford algebra, which we may suppress for notational simplicity and re-introduce when needed. The action consists of a first-order Dirac-type differential operator $\hat{H} \equiv \gamma^\mu D_\mu + m$ acting on spinors.  As is standard procedure~\cite{Vassilevich:2003xt,DeBerredoPeixoto:2001qm}, we can define the determinant of a Dirac operator $\hat{H}$ as the square root of the determinant of $\hat{H}\hat{H}^\dagger$:
\begin{equation}
	\log\det\hat{H} = \log\det\hat{H}^\dagger = \frac{1}{2}\log\det\hat{H}\hat{H}^\dagger~.
\end{equation}
It is therefore sufficient to compute the heat kernel of $\hat{H}\hat{H}^\dagger$.  

Let us now assume that our spacetime is even-dimensional and has Euclidean signature, in which case our gamma matrices are Hermitian, $\gamma_\mu^\dagger = \gamma_\mu$.  With this choice, the operator $\gamma^\mu D_\mu$ is anti-Hermitian, $\left(\gamma^\mu D_\mu\right)^\dagger = -\gamma^\mu D_\mu$, and hence we find that the relevant second-order differential operator acting on $\psi$ is
\begin{equation}
	\Lambda = \hat{H}\hat{H}^\dagger = -\gamma^\mu \gamma^\nu D_\mu D_\nu + m^2~.
\end{equation}
The covariant derivative acts on $\psi$ by
\begin{equation}
	D_\mu \psi = \partial_\mu \psi + \frac{1}{4}\gamma_{ab}\omega^{ab}_\mu \psi~,
\end{equation}
for the spin connection $\omega^{ab}_\mu$, where  $\mu,\nu,...$ are curved space indices, $a,b,...$ are flat space indices, $\gamma^\mu$ is shorthand for $\gamma^a e^\mu_a$, and $e^\mu_a$ is the vierbein.  Gamma matrix commutation relations give
\begin{equation}
	[D_\mu,D_\nu]\psi = \frac{1}{4}\gamma^{ab} R_{\mu\nu a b}\psi = \frac{1}{4}\gamma^{\rho\sigma} R_{\mu\nu\rho\sigma}\psi~.
\label{eq:commutator_spinor}
\end{equation}
By breaking up $\gamma^\mu \gamma^\nu D_\mu D_\nu$ into its symmetric and antisymmetric parts and using (\ref{eq:commutator_spinor}), we find that
\begin{equation}
\label{eq:lambda_spinor}
\begin{aligned}
	\bar{\psi}\Lambda\psi &= \bar{\psi}\left(-\frac{1}{2}\gamma^\rho\gamma^\sigma \{D_\rho,D_\sigma\} - \frac{1}{2}\gamma^{\rho\sigma}[D_\rho,D_\sigma] +m^2\right)\psi \\
	&= \bar{\psi}\left(-\square - \frac{1}{8}\gamma^{\mu\nu}\gamma^{\rho\sigma} R_{\mu\nu\rho\sigma} + m^2\right)\psi \\
	&=  \bar{\psi}\left(-\square + \frac{1}{4}R + m^2\right)\psi~,
\end{aligned}
\end{equation}
where equality in the last line comes from using gamma matrix commutation relations and the Bianchi identity $R_{\mu[\nu\rho\sigma]} = 0$.

$\Lambda$ defined in (\ref{eq:lambda_spinor}) is of the Laplace-type form required in (\ref{eq:lambda_def}), and we identify $I$, $\omega_\mu$, $E$, and $\Omega_{\mu\nu}$ as
\begin{equation}
	I = \mathbb{I}_4~, \quad \omega_\mu = 0~, \quad E = -m^2\mathbb{I}_4~, \quad \Omega_{\mu\nu} = \frac{1}{4}\gamma^{\rho\sigma} R_{\mu\nu\rho\sigma}~.
\end{equation}
We can use gamma matrix identities to compute the traces of $I$, $E$, $E^2$, and $\Omega_{\mu\nu}\Omega^{\mu\nu}$ needed for our heat kernel coefficients.  The result is:
\begin{equation}
\label{eq:a2n_spinor}
\begin{aligned}
(4\pi)^{2} a_0(x) &= -4~,\\
(4\pi)^{\frac{D}{2}}a_2(x) &= 4m^2~, \\
(4\pi)^{\frac{D}{2}}a_4(x) &= -\bigg{[}2 m^4 + \frac{1}{360}\left(-7 R_{\mu\nu\rho\sigma}R^{\mu\nu\rho\sigma}-8R_{\mu\nu}R^{\mu\nu}\right) \bigg{]}~.
\end{aligned}
\end{equation}
The overall minus sign on each of these heat kernel coefficients is put in by hand to account for fermion statistics. We also note that this derivation assumed that $\psi$ was a Dirac spinor.  Weyl and Majorana spinors have half the degrees of freedom of Dirac spinors, and so we must divide these results by two if we want the heat kernel coefficients for Majorana or Weyl spinors.

This derivation was done in a Euclidean spacetime in order to take advantage of Hermitian gamma matrices.  For Lorentzian spacetimes the spinor conjugation includes $\gamma^0$, which has the effect of changing the boundary conditions on the conjugate spinor.  We are considering manifolds without boundary, so this change in boundary conditions is irrelevant and our results naturally generalize to Lorentzian spacetimes as well~\cite{Vassilevich:2003xt}.

%%%%%%%%%%%%%%%%%%%%%%%%%%%%%%%%%%%%%%%%%%%%%%%%%%%%%%%%%%%%%%

\subsection{$\mathcal{N}=2$ Multiplet Heat Kernels}
\label{subsec:heatkernel_n=2}

We will now compute the heat kernels of the quadratic fluctuations of the $\mathcal{N}=2$ multiplet field content.  The heat kernel coefficient formulae (\ref{eq:a2n}) are strictly in terms of local invariants constructed from the background fields, so there are no issues in using the classical equations of motion for the background fields to simplify these expressions.  We will freely make use of the Ricci-flat Einstein equation
\begin{equation}
\label{eq:einstein_r=0}
	R_{\mu\nu} = \frac{1}{2}F_{\mu\rho}F_\nu^{~\rho} -\frac{1}{4}g_{\mu\nu}F_{\rho\sigma}F^{\rho\sigma}~,
\end{equation}
as well as the Schouten identity (given in equation (4.47) of~\cite{Freedman:2012zz})
\begin{equation}
\label{eq:schouten}
	F_{\mu\rho}\widetilde{F}_\nu^{~\rho} = \widetilde{F}_{\mu\rho}F_\nu^{~\rho} = \frac{1}{4}g_{\mu\nu}F_{\rho\sigma}\widetilde{F}^{\rho\sigma}~.
\end{equation}
We will also use Maxwell's equations and the Bianchi identity in the form
\begin{equation}
\label{eq:maxwell}
	D^\mu F_{\mu\nu} = 0~,\quad D^\mu\widetilde{F}_{\mu\nu} = 0~,
\end{equation}
and the gravitational Bianchi identity
\begin{equation}
\label{eq:bianchi}
	R_{\mu[\nu\rho\sigma]} = 0~.
\end{equation}
In addition, we will make extensive use of gamma matrix technology in 4D, using the conventions from~\cite{Freedman:2012zz}.  In particular, the identity
\begin{equation}
	\gamma_{\mu\nu\rho\sigma} = -i \gamma_5 \varepsilon_{\mu\nu\rho\sigma}
\label{eq:gamma5ep}
\end{equation}
will be very useful.  Lastly, we can ignore total derivative terms in heat kernel coefficients and so we freely integrate by parts.  For example, we find that (up to a total derivative)
\begin{equation}\begin{aligned}
	(D_\rho F_{\mu\nu})(D^\rho F^{\mu\nu}) &= 2 (D_\rho F_{\mu\nu})(D^\nu F^{\mu\rho}) \\
	&= -2 (D^\nu D_\rho F_{\mu\nu})F^{\mu\rho} \\
	&= -2([D_\nu,D_\rho]F^{\mu\nu})F_\mu^{~\rho}~,
\end{aligned}\end{equation}
where equality comes from the Bianchi identity, integration by parts, and Maxwell's equations, respectively.  The covariant derivative commutator acting on a rank-2 tensor then gives
\begin{equation}
\label{eq:dfdf}
	(D_\rho F_{\mu\nu})(D^\rho F^{\mu\nu}) = -2 R_{\mu\nu}F^{\mu\rho}F^\nu_{~\rho} + R_{\mu\nu\rho\sigma}F^{\mu\nu}F^{\rho\sigma}~.
\end{equation}
We note that (\ref{eq:dfdf}) also holds if we replace $F_{\mu\nu}$ with the dual field strength $\widetilde{F}_{\mu\nu}$.

%%%%%%%%%%%%%%%%%%%%%%%%%%%%%%%%%%%%%%%%%%%%%%%%%%%%%%%%%%%%%%

\subsubsection{Hyper Multiplet}
\label{subsubsec:heatkernel_n=2_hypermultiplet}

A single $\mathcal{N}=2$ hyper multiplet contains two Majorana fermions and four real scalars.  The scalars are minimally coupled to gravity and massless, so we can use the free scalar result (\ref{eq:a2n_scalar}) with $m = 0$:
\begin{equation}
\label{eq:hk_hyper_b}
\begin{aligned}
	(4\pi)^2 a_0^{\text{H},b}(x) &= 4~, \\
	(4\pi)^2 a_2^{\text{H},b}(x) &= 0~, \\
	(4\pi)^2 a_4^{\text{H},b}(x) &= \frac{1}{45}\left(R_{\mu\nu\rho\sigma}R^{\mu\nu\rho\sigma}-R_{\mu\nu}R^{\mu\nu}\right)~.
\end{aligned}
\end{equation}

The Lagrangian for the hyper fermions (\ref{eq:ed}) mixes left-handed and right-handed fermions.  We want to put this Lagrangian in the form of a diagonal Dirac-type Lagrangian to use the procedure outlined earlier for fermionic heat kernels.  We define the spinor
\begin{equation}
	\psi_A \equiv P_R \zeta_A + P_L \zeta^A~,
\end{equation}
where $P_L = \frac{1}{2}(1 + \gamma_5)$, $P_R = \frac{1}{2}(1 - \gamma_5)$. The hyper fermion Lagrangian can then be rewritten as
\begin{equation}
\label{eq:lhyper_dirac}
	\mathcal{L}_\text{hyper} = \bar{\psi}_A\left( - \gamma^\mu D_\mu \delta_{AB} + \frac{1}{4} F_{\mu\nu}\gamma^{\mu\nu} \epsilon_{AB}\right)\psi_B~.
\end{equation}
Though we have lost information about the chirality of the spinors, this Lagrangian is now in the form of (\ref{eq:l_h}).  That is, we can express the Lagrangian as
\begin{equation}
	\mathcal{L}_\text{hyper} = \bar{\psi}_A \hat{H}_{AB} \psi_B~,
\end{equation}
where $\hat{H}_{AB}$ is a Dirac operator acting on the spinors $\psi_A$ by
\begin{equation}
	\hat{H}_{AB} = - \gamma^\mu D_\mu \delta_{AB} + \frac{1}{4} F_{\mu\nu}\gamma^{\mu\nu} \epsilon_{AB}~.
\end{equation}
As with the free spinor field we now continue to Euclidean space, giving us Hermitian gamma matrices $\gamma_\mu^\dagger = \gamma_\mu$ and the spinor conjugate $\bar{\psi}_A=\psi^\dagger_A$.  We can also choose the background field $F^{\mu\nu}$ to be real.  With these conventions, the Hermitian conjugate of $\hat{H}_{AB}$ is
\begin{equation}
	\hat{H}_{AB}^\dagger = \gamma^\mu D_\mu \delta_{AB} + \frac{1}{4} F_{\mu\nu}\gamma^{\mu\nu} \epsilon_{AB}~.
\end{equation}
The relevant Laplace-type operator that we will compute the heat kernel of is 
\begin{equation}\begin{aligned}
	\Lambda_{AB} &= \hat{H}_{AC}\hat{H}_{CB}^\dagger \\
	&=\left( - \gamma^\mu D_\mu  \delta_{AC} + \frac{1}{4} F_{\mu\nu}\gamma^{\mu\nu} \epsilon_{AC}\right) \left( \gamma^\mu D_\mu  \delta_{CB} + \frac{1}{4} F_{\mu\nu}\gamma^{\mu\nu} \epsilon_{CB}\right) \\
	&= -\left(\gamma^\mu\gamma^\nu D_\mu D_\nu	\delta_{AB} -\frac{1}{4} F_{\mu\nu}\gamma^{\mu\nu} \gamma^\rho D_\rho  \epsilon_{AB} + \frac{1}{4} \gamma^\rho D_\rho  F_{\mu\nu}\gamma^{\mu\nu} \epsilon_{AB} + \frac{1}{16}F_{\mu\nu} F_{\rho\sigma}\gamma^{\mu\nu}\gamma^{\rho\sigma} \delta_{AB}\right) \\
	&= -\bigg{[}\square \delta_{AB} + F_{\rho\mu}\gamma^\mu D^\rho \epsilon_{AB} + \left(\frac{1}{8}F_{\mu\nu}\widetilde{F}^{\mu\nu}\gamma_5 - \frac{1}{8}F_{\mu\nu}F^{\mu\nu} \right)\delta_{AB}\bigg{]}~,
\label{eq:hypersquared}
\end{aligned}\end{equation}
where equality in the last line comes from using (\ref{eq:lambda_spinor}) and (\ref{eq:gamma5ep}), as well as noting that
\begin{equation}
\label{eq:df0}
	\gamma^\rho (D_\rho F_{\mu\nu})\gamma^{\mu\nu} = (D_\rho  F_{\mu\nu})\gamma^{\mu\nu}\gamma^\rho = 0~,
\end{equation}
by the Maxwell-Bianchi equations (\ref{eq:maxwell}).

From the form of $\Lambda$ in (\ref{eq:hypersquared}) we identify the matrices $I$, $\omega_\mu$, and $P$ as
\begin{equation}
\label{eq:hyper_iwp}
	I_{AB} = \mathbb{I}_4 \delta_{AB}~, \quad (\omega_\mu)_{AB} = \left(\frac{1}{2}F_{\mu\nu}\gamma^\nu\right)\epsilon_{AB}~, \quad	P_{AB} = \frac{1}{8}\left(F_{\mu\nu}\widetilde{F}^{\mu\nu}\gamma_5 - F_{\mu\nu}F^{\mu\nu} \right)\delta_{AB}~.
\end{equation}
By Maxwell's equations $\big{(}D^\mu (\omega_\mu)_{AB}\big{)} = 0$ so $E$ becomes
\begin{equation}
	E_{AB} = P_{AB} - (\omega_\mu)_{AC} (\omega^\mu)_{CB} = \frac{1}{8}\left(F_{\mu\nu}\widetilde{F}^{\mu\nu}\gamma_5 + F_{\mu\nu}F^{\mu\nu} \right)\delta_{AB}~.
\end{equation}
The curvature $\Omega_{\mu\nu}$ corresponding to the effective covariant derivative $\mathcal{D}_\mu = D_\mu + \omega_\mu$ is 
\begin{equation}\begin{aligned}
	(\Omega_{\mu\nu})_{AB} &= [(\mathcal{D}_\mu)_{AC},(\mathcal{D}_\nu)_{CB}] \\
	&= [D_\mu,D_\nu]\delta_{AB} + \big{(}D_{\mu}(\omega_{\nu})_{AB}\big{)} -  \big{(}D_{\nu}(\omega_{\mu})_{AB}\big{)} + [(\omega_\mu)_{AC},(\omega_\nu)_{CB}]
\end{aligned}\end{equation}
We can use our expressions for $[D_\mu,D_\nu]$ and $\omega_\mu$ from (\ref{eq:commutator_spinor}) and (\ref{eq:hyper_iwp}), respectively, giving
\begin{equation}
	(\Omega_{\mu\nu})_{AB} = \left( \frac{1}{4}R_{\mu\nu\rho\sigma} - \frac{1}{2}F_{\mu\rho}F_{\nu\sigma}\right)\gamma^{\rho\sigma}\,\delta_{AB} + \left(-\frac{1}{2}\gamma^\rho D_\rho F_{\mu\nu}\right)\epsilon_{AB}~.
\end{equation}
We can now compute all of the traces necessary for the Seeley-DeWitt coefficients.  These traces are tedious but straightforward to compute, so we will simply quote the results, noting that we use (\ref{eq:einstein_r=0}) and (\ref{eq:dfdf}) to simplify when possible.
\begin{equation}\begin{aligned}
	\text{Tr }I &= 8~, \\
	\text{Tr }E &= F_{\mu\nu}F^{\mu\nu}~, \\
	\text{Tr }E^2 &=\frac{1}{8}(F_{\mu\nu}F^{\mu\nu})^2 + \frac{1}{8}(F_{\mu\nu}\widetilde{F}^{\mu\nu})^2~, \\
	\text{Tr }\Omega_{\mu\nu}\Omega^{\mu\nu} &= -R_{\mu\nu\rho\sigma}R^{\mu\nu\rho\sigma} +16 R_{\mu\nu}R^{\mu\nu} - \frac{3}{2}(F_{\mu\nu}F^{\mu\nu})^2~.
\label{eq:hypertrace1}
\end{aligned}\end{equation}

We can use these quantities with (\ref{eq:a2n}) to calculate the Seeley-DeWitt coefficients for the hyper fermions, making sure to add an overall factor of $-1/2$ to account for fermion statistics and the Majorana condition. The result is
\begin{equation}\begin{aligned}
\label{eq:hk_hyper_f}
(4\pi)^2 a_0^{\text{H},f}(x) &= -4~, \\
(4\pi)^2 a_2^{\text{H},f}(x) &= -\frac{1}{2}F_{\mu\nu}F^{\mu\nu}~, \\
(4\pi)^2 a_4^{\text{H},f}(x) &= -\frac{1}{360}\left( -7R_{\mu\nu\rho\sigma}R^{\mu\nu\rho\sigma} + 232R_{\mu\nu}R^{\mu\nu} -\frac{45}{4}(F_{\mu\nu}F^{\mu\nu})^2 + \frac{45}{4}(F_{\mu\nu}\widetilde{F}^{\mu\nu})^2\right)~.
\end{aligned}\end{equation} 
Adding up the bosonic part (\ref{eq:hk_hyper_b}) and the fermionic part (\ref{eq:hk_hyper_f}), the full hyper multiplet heat kernel coefficients are
\begin{equation}\begin{aligned}
\label{eq:hk_hyper}
(4\pi)^2 a_0^\text{H}(x) &= 0~, \\
(4\pi)^2 a_2^\text{H}(x) &= -\frac{1}{2}F_{\mu\nu}F^{\mu\nu}~, \\
(4\pi)^2 a_4^\text{H}(x) &= \frac{1}{24}\left(R_{\mu\nu\rho\sigma}R^{\mu\nu\rho\sigma}-16 R_{\mu\nu}R^{\mu\nu}+\frac{3}{4}(F_{\mu\nu}F^{\mu\nu})^2 -\frac{3}{4}(F_{\mu\nu}\widetilde{F}^{\mu\nu})^2\right)~.
\end{aligned}\end{equation}
The $a_0(x)$ coefficient vanishes because any full multiplet has an equal number of bosonic and fermionic degrees of freedom.

%%%%%%%%%%%%%%%%%%%%%%%%%%%%%%%%%%%%%%%%%%%%%%%%%%%%%%%%%%%%%%

\subsubsection{Vector Multiplet}
\label{subsubsec:heatkernel_n=2_vectormultiplet}

The $\mathcal{N}=2$ vector multiplet consists of one vector field, two gauginos, and one complex scalar.  The gauginos are massless Majorana fermions that couple minimally to gravity, and thus we can use (\ref{eq:a2n_spinor}) to find the vector multiplet fermionic heat kernel coefficients
\begin{equation}\begin{aligned}
\label{eq:hk_vector_f}
	(4\pi)^2 a_0^{\text{V},f}(x) &= -4~, \\
	(4\pi)^2 a_2^{\text{V},f}(x) &= 0~, \\
	(4\pi)^2 a_4^{\text{V},f}(x) &= -\frac{1}{360}\left(-7R_{\mu\nu\rho\sigma}R^{\mu\nu\rho\sigma}-8R_{\mu\nu}R^{\mu\nu}\right)~.
\end{aligned}\end{equation}

The equations of motion for the bosonic content of the vector multiplet are given in (\ref{eq:do}).  We split the complex scalar $z$ into its real and imaginary parts by
\begin{equation}
	z = x - i y~,
\end{equation}
where $x$ is a real pseudoscalar field and $y$ is a real scalar field.  The bosonic action consistent with the equations of motion (\ref{eq:do}) is
\begin{equation}\begin{aligned}
	\mathcal{L}_b &= -\frac{1}{8}f_{\mu\nu}f^{\mu\nu} - \frac{1}{4}(D_\mu y)(D^\mu y) + \frac{1}{4}y f_{\mu\nu}F^{\mu\nu} - \frac{1}{16}y^2 F_{\mu\nu}F^{\mu\nu} \\
	&- \frac{1}{4}(D_\mu x)(D^\mu x) + \frac{i}{4}x f_{\mu\nu}\widetilde{F}^{\mu\nu} + \frac{1}{16}x^2 \widetilde{F}_{\mu\nu}\widetilde{F}^{\mu\nu} - \frac{i}{8}x y F_{\mu\nu}\widetilde{F}^{\mu\nu}~,
\end{aligned}\end{equation}
where $f_{\mu\nu} = D_\mu a_\nu - D_\nu a_\mu$ is the fluctuation about the background field strength $F_{\mu\nu}$.  As a consistency check we note that on AdS$_2\times S^2$ (where $F_{\mu\nu}\widetilde{F}^{\mu\nu} = 0$) this action is consistent with equation (6.4) of~\cite{Sen:2011ba}.

We choose the Lorenz gauge $D_\mu a^\mu = 0$ by adding a gauge-fixing term to the Lagrangian
\begin{equation}
	\mathcal{L}_{\text{g.f.}} = -\frac{1}{4}(D_\mu a^\mu)^2~.
\end{equation}
This gauge-fixing will introduce two anti-commuting scalar ghosts that will contribute to the heat kernel with an overall minus sign.  We denote $\{\phi_m\} = \{y,x,a_\mu\}$ to be the bosonic field fluctuations.  Then, we can use Maxwell's equations and the Bianchi identity to rewrite our action in the Hermitian form required in (\ref{eq:lambda_def}), up to a total derivative, as
\begin{equation}
	\mathcal{S} = -\frac{1}{4}\int d^4x\,\sqrt{-g} \phi_n \Lambda^n_m \phi_m~,
\end{equation}
where
\begin{equation}\begin{aligned}
\label{eq:lambda_vector}
	-\phi_n \Lambda^n_m \phi_m &= a_\mu\left(\square g^{\mu\nu} - R^{\mu\nu}\right)a_\nu + y\left(\square - \frac{1}{4}F_{\mu\nu}F^{\mu\nu}\right)y + x\left(\square + \frac{1}{4}\widetilde{F}_{\mu\nu}\widetilde{F}^{\mu\nu}\right)x \\
	&+y\left(F^{\rho\nu}D_\rho\right)a_\nu + a_\mu\left(F^{\mu\rho}D_\rho\right)y +x\left(i\widetilde{F}^{\rho\nu}D_\rho\right)a_\nu + a_\mu\left(i\widetilde{F}^{\mu\rho}D_\rho\right)x \\
	& + y\left(-\frac{i}{4}F_{\mu\nu}\widetilde{F}^{\mu\nu}\right)x + x\left(-\frac{i}{4}F_{\mu\nu}\widetilde{F}^{\mu\nu}\right)y~.
\end{aligned}\end{equation}
From (\ref{eq:lambda_vector}) we can read off the matrices $P$ and $\omega_\rho$. And, since all of the terms in $\omega_\rho$ depend on $F_{\mu\rho}$ or $\widetilde{F}_{\mu\rho}$, $(D^\rho \omega_\rho) = 0$ due to Maxwell's equations and the Bianchi identities.  $E$ thus becomes
\begin{equation}
\label{eq:vector_e}
\phi_n E^n_m \phi^m = \phi_n \left(P - \omega^\rho \omega_\rho\right)^n_m \phi^m = a_\mu\left(-R^{\mu\nu} + \frac{1}{4}F^{\mu\rho}F^\nu_{~\rho}-\frac{1}{4}\widetilde{F}^{\mu\rho}\widetilde{F}^\nu_{~\rho}\right)a_\nu~.
\end{equation}
The lack of any terms involving $x$ or $y$ in (\ref{eq:vector_e}) was not a priori obvious but a consequence of how terms in the action that coupled $x$ and $y$ to the background conspired to cancel.  

There are six off-shell bosonic degrees of freedom for the fields $\{\phi_n\}$: four from the vector $a_\mu$, and two from the scalars $x$ and $y$, giving $\text{Tr }I = 6$.  From (\ref{eq:vector_e}) we compute the traces
\begin{equation}\begin{aligned}
	\text{Tr }E &= \frac{1}{4}\left(F_{\mu\nu}F^{\mu\nu} - \widetilde{F}_{\mu\nu}\widetilde{F}^{\mu\nu}\right)~, \\
	\text{Tr }E^2 &= R_{\mu\nu}R^{\mu\nu} - \frac{1}{2}R_{\mu\nu}F^{\mu\rho}F^\nu_{~\rho} + \frac{1}{2}R_{\mu\nu}\widetilde{F}^{\mu\rho}\widetilde{F}^\nu_{~\rho} \\
	&\quad + \frac{1}{16}(F^{\mu\rho}F_{\nu\rho})(F_{\mu\sigma}F^{\nu\sigma}) + \frac{1}{16}(\widetilde{F}^{\mu\rho}\widetilde{F}_{\nu\rho})(\widetilde{F}_{\mu\sigma}\widetilde{F}^{\nu\sigma}) - \frac{1}{8}(F^{\mu\rho}F_{\nu\rho})(\widetilde{F}_{\mu\sigma}\widetilde{F}^{\nu\sigma})~.
\end{aligned}\end{equation}

In order to compute the curvature we expand the commutator in (\ref{eq:omeg})
\begin{equation}
	(\Omega_{\rho\sigma})^n_m = [D_\rho,D_\sigma]^n_m + (D_{[\rho}\omega_{\sigma]})^n_m + [\omega_\rho,\omega_\sigma]^n_m~.
\label{eq:om2}
\end{equation}
The covariant derivative commutes when acting on scalars, but not for vectors, and so the first term in (\ref{eq:om2}) is
\begin{equation}
	\phi_n[D_\rho,D_\sigma]^n_m \phi^m = a_\mu [D_\rho,D_\sigma]a^\mu = a_\mu\left(R^\mu_{~\nu\rho\sigma}\right)a^\nu~.
\end{equation}
The second term in (\ref{eq:om2}) is calculated by applying the covariant derivative to $\omega_\mu$.  Using the Maxwell-Bianchi equations to simplify we find that
\begin{equation}\begin{aligned}
	\phi_n( D_{[\rho}\omega_{\sigma]})^n_m\phi^m &= y\left(-\frac{1}{2}(D_\nu F_{\rho\sigma})\right)a^\nu + a_\mu\left(\frac{1}{2}(D^\mu F_{\rho\sigma})\right) y \\
	&\quad + x\left(-\frac{i}{2}(D_\nu \widetilde{F}_{\rho\sigma})\right)a^\nu + a_\mu\left(\frac{i}{2}(D^\mu \widetilde{F}_{\rho\sigma})\right) y~.
\end{aligned}\end{equation}
The last term in (\ref{eq:om2}) is the product of $\omega_\rho$ and $\omega_\sigma$, antisymmetrized in $\rho$ and $\sigma$:
\begin{equation}
	\phi_n [\omega_\rho,\omega_\sigma]^n_m\phi^m = a_\mu \left(\frac{1}{4} F^\mu_{~[\rho}F_{\sigma]\nu} - \frac{1}{4} \widetilde{F}^\mu_{~[\rho}\widetilde{F}_{\sigma]\nu}\right)a^\nu~.
\end{equation}
Adding all of these components up, we find that
\begin{equation}\begin{aligned}
	\phi_n (\Omega_{\rho\sigma})^n_m\phi^m &= a_\mu \left(R^\mu_{~\nu\rho\sigma} + \frac{1}{4} F^\mu_{~[\rho}F_{\sigma]\nu} - \frac{1}{4} \widetilde{F}^\mu_{~[\rho}\widetilde{F}_{\sigma]\nu}\right)a^\nu \\
	&\quad + y\left(-\frac{1}{2}(D_\nu F_{\rho\sigma})\right)a^\nu + a_\mu\left(\frac{1}{2}(D^\mu F_{\rho\sigma})\right) y \\
	&\quad + x\left(-\frac{i}{2}(D_\nu \widetilde{F}_{\rho\sigma})\right)a^\nu + a_\mu\left(\frac{i}{2}(D^\mu \widetilde{F}_{\rho\sigma})\right) y~.
\end{aligned}\end{equation}
Now that we have all of the components of $\Omega_{\rho\sigma}$, it is straightforward to trace over $\Omega_{\rho\sigma}\Omega^{\rho\sigma}$.  We will also simplify by using the Maxwell-Bianchi equations and (\ref{eq:dfdf}).  The result, up to a total derivative, is
\begin{equation}\begin{aligned}
	\text{Tr }\Omega_{\rho\sigma}\Omega^{\rho\sigma}
	&= -R_{\mu\nu\rho\sigma}R^{\mu\nu\rho\sigma} + R_{\mu\nu}\left(F^{\mu\rho}F^\nu_{~\rho} - \widetilde{F}^{\mu\rho}\widetilde{F}^\nu_{~\rho}\right) \\
	&\quad + \frac{1}{8}\left( (F^{\mu\rho}F_{\nu\rho})(F_{\mu\sigma}F^{\nu\sigma}) + (\widetilde{F}^{\mu\rho}\widetilde{F}_{\nu\rho})(\widetilde{F}_{\mu\sigma}\widetilde{F}^{\nu\sigma}) - (F_{\mu\nu}F^{\mu\nu})^2 - (\widetilde{F}_{\mu\nu}\widetilde{F}^{\mu\nu})^2 \right) \\
	&\quad+\frac{1}{4}\left( (F_{\mu\nu}\widetilde{F}^{\mu\nu})^2 - (F^{\mu\rho}\widetilde{F}_{\nu\rho})(\widetilde{F}_{\mu\sigma}F^{\nu\sigma}) \right)~.
\end{aligned}\end{equation}

We now have all of the traces needed to calculate the Seeley-DeWitt coefficients for the bosonic fields.  However, our gauge-fixing also introduced two scalar ghosts into our system.  These ghosts do not interact with any of the bosonic fields and so their corresponding heat kernels are those for two minimally coupled scalars (\ref{eq:a2n_scalar}).  If we insert our traces into the coefficient formulas in (\ref{eq:a2n}) and subtract off the ghost contribution, we find that:
\begin{equation}
\label{eq:hk_vector_b}
\begin{aligned}
	(4\pi)^2 a_0^{\text{V},b}(x) &= 4~, \\
	(4\pi)^2 a_2^{\text{V},b}(x) &= \frac{1}{4}\left(F_{\mu\nu}F^{\mu\nu} - \widetilde{F}_{\mu\nu}\widetilde{F}^{\mu\nu}\right)~, \\
	(4\pi)^2 a_4^{\text{V},b}(x) &= \frac{1}{180}\bigg{[} -11 R_{\mu\nu\rho\sigma}R^{\mu\nu\rho\sigma}+86R_{\mu\nu}R^{\mu\nu} -30R_{\mu\nu}\left(F^{\mu\rho}F^\nu_{~\rho} - \widetilde{F}^{\mu\rho}\widetilde{F}^\nu_{~\rho}\right)\\
	&\quad\quad\quad +\frac{15}{2}\left(F^{\mu\rho}F_{\nu\rho}\right)\left(F_{\mu\sigma}F^{\nu\sigma}\right) + \frac{15}{2}\left(\widetilde{F}^{\mu\rho}\widetilde{F}_{\nu\rho}\right)\left(\widetilde{F}_{\mu\sigma}\widetilde{F}^{\nu\sigma}\right) \\
	&\quad\quad\quad - \frac{15}{8}\left(F_{\mu\nu}F^{\mu\nu}\right)^2 - \frac{15}{8}\left(\widetilde{F}_{\mu\nu}\widetilde{F}^{\mu\nu}\right)^2 \\
	&\quad\quad\quad + \frac{15}{4}\left( F_{\mu\nu}\widetilde{F}^{\mu\nu}\right)^2 -\frac{45}{4}\left(F^{\mu\rho}F_{\mu\rho}\right)\left(\widetilde{F}_{\mu\sigma}\widetilde{F}^{\mu\sigma}\right) - \frac{15}{4}\left(F^{\mu\rho}\widetilde{F}_{\nu\rho}\right)\left(\widetilde{F}_{\mu\sigma}F^{\nu\sigma}\right) \bigg{]}~.
\end{aligned}\end{equation}

Adding up the fermionic (\ref{eq:hk_vector_f}) and bosonic (\ref{eq:hk_vector_b}) contributions and using the Schouten identity (\ref{eq:schouten}) to simplify, the full vector multiplet heat kernel is
\begin{equation}
\label{eq:hk_vector}
\begin{aligned}
	(4\pi)^2 a_0^{\text{V}}(x) &= 0~, \\
	(4\pi)^2 a_2^{\text{V}}(x) &= \frac{1}{4}\left(F_{\mu\nu}F^{\mu\nu} - \widetilde{F}_{\mu\nu}\widetilde{F}^{\mu\nu}\right)~, \\
	(4\pi)^2 a_4^{\text{V}}(x) &= \frac{1}{24}\bigg{[} - R_{\mu\nu\rho\sigma}R^{\mu\nu\rho\sigma}+12R_{\mu\nu}R^{\mu\nu} -4R_{\mu\nu}\left(F^{\mu\rho}F^\nu_{~\rho} - \widetilde{F}^{\mu\rho}\widetilde{F}^\nu_{~\rho}\right)\\
	&\quad\quad\quad +\left(F^{\mu\rho}F_{\nu\rho}\right)\left(F_{\mu\sigma}F^{\nu\sigma}\right) + \left(\widetilde{F}^{\mu\rho}\widetilde{F}_{\nu\rho}\right)\left(\widetilde{F}_{\mu\sigma}\widetilde{F}^{\nu\sigma}\right) \\
	&\quad\quad\quad - \frac{1}{4}\left(F_{\mu\nu}F^{\mu\nu}\right)^2 - \frac{1}{4}\left(\widetilde{F}_{\mu\nu}\widetilde{F}^{\mu\nu}\right)^2\bigg{]}~.
\end{aligned}\end{equation}

Our result for $a_2^V(x)$ disagrees with~\cite{Keeler:2014bra} in the special case of BPS black holes.  However, $a_2(x)$ determines the quadratic divergences and encodes the renormalization of the Newton constant.  These quadratic divergences are scheme-dependent and unphysical.  We will record our results for $a_2(x)$ in our heat kernel regularization scheme for the sake of completion.

%%%%%%%%%%%%%%%%%%%%%%%%%%%%%%%%%%%%%%%%%%%%%%%%%%%%%%%%%%%%%%

\subsubsection{Gravity Multiplet: Fermions}
\label{subsubsec:heatkernel_n=2_gravitymultiplet_f}

The gravity multiplet consists of the graviton, two Majorana gravitini, and the graviphoton.  We rewrite the Lagrangian for these gravitini (\ref{eq:eb}) by using (\ref{eq:gamma5ep}) to express $\gamma^{\mu\nu\rho\sigma}$ in terms of $\gamma_5$ and the Levi-Civita symbol, resulting in
\begin{equation}
\label{eq:lgravitini}
	\mathcal{L}_\text{gravitini} = -\frac{1}{2\kappa^2}\bar{\Psi}_{A\mu}\gamma^{\mu\nu\rho}D_\nu \Psi_{A\rho} + \frac{1}{4\kappa^2}\bar{\Psi}_{A\mu}\left(F^{\mu\nu} + \gamma_5 \widetilde{F}^{\mu\nu}\right)\epsilon_{AB}\Psi_{B\nu}~,
\end{equation}
where $A,B=1,2$ enumerates the two gravitini species. The covariant derivative acts on the gravitino field $\Psi^\rho_A$ by
\begin{equation}
	D_{\mu}\Psi^\rho_A = \partial_\mu \Psi^\rho_A + \frac{1}{4}\gamma_{ab} \omega^{ab}_\mu \Psi^\rho_A +  \Gamma^\rho_{\mu\nu}\Psi^\nu_A~,
\end{equation}
for the spin connection $\omega^{ab}_\mu$ and the Levi-Civita connection $\Gamma^{\rho}_{\mu\nu}$. The commutator $[D_\mu,D_\nu]$ acting on $\Psi^\rho_A$ will be the sum of the spin and Riemann commutators
\begin{equation}
	\Bar{\Psi}_{A\rho}[D_\mu,D_\nu]\Psi^\rho_A = \Bar{\Psi}_{A\rho}\left(\frac{1}{4}g_{\rho\sigma}\gamma^{\alpha\beta} R_{\mu\nu\alpha\beta} + R_{\mu\nu}^{~~\rho\sigma}\right)\delta_{AB}\Psi_{B \sigma}~.
\label{eq:commutator_gravitino}
\end{equation}
Tbe gravitini Lagrangian (\ref{eq:lgravitini}) is invariant under the SUSY transformation
\begin{equation}
	\delta\Psi_{A\mu} = \left(\delta_{AB} D_\mu - \frac{1}{8}\epsilon_{AB}\gamma^{\rho\sigma}F_{\rho\sigma}\gamma_\mu \right) \epsilon_B~,
\end{equation}
for a spinor $\epsilon_B$.  This SUSY transformation acts as a gauge symmetry.

We need the kinetic term of the gravitini to be in Dirac form in order for it to square to a minimal operator.  We use the procedure outlined in~\cite{Nielsen:1978ex} and gauge-fix our action in such a way that, when paired with a suitable corresponding field redefinition, the kinetic term becomes Dirac-type.  In particular, we choose the harmonic gauge for our gravitini $\gamma^\mu \Psi_{A\mu} = 0$ by adding the gauge-fixing term
\begin{equation}
	\mathcal{L}_{\text{g.f.}} = \frac{1}{4\kappa^2}(\bar{\Psi}_{A\mu}\gamma^\mu)\gamma^\nu D_\nu(\gamma^\rho \Psi_{A\rho})~.
\label{eq:gravitino_gauge}
\end{equation}
Then, we consider the field redefinition
\begin{equation}
\label{eq:gravitino_redefinition}
	\Phi_{A\mu} = \Psi_{A\mu} - \frac{1}{2}\gamma_{\mu}\gamma^\nu \Psi_{A\nu}~.
\end{equation}
Using gamma matrix identities and (\ref{eq:gamma5ep}), it is easily verified that
\begin{equation}\begin{aligned}
	\bar{\Phi}_{A\mu} \gamma^\nu D_\nu \Phi_A^\mu &= \bar{\Psi}_{A\mu} \left( \gamma^{\mu\nu\rho}D_\nu - \frac{1}{2} \gamma^\mu \gamma^\nu \gamma^\rho D_\nu \right) \Psi_{A\rho}~, \\
	\bar{\Phi}_{A\mu} F^{\mu\nu} \Phi_{B\nu} &= \frac{1}{2}\bar{\Psi}_{A\mu}\left(F^{\mu\nu} + \gamma_5 \widetilde{F}^{\mu\nu} + \frac{1}{2}\gamma^{\rho\sigma}F_{\rho\sigma}g^{\mu\nu}\right)\Psi_{B\nu}~,\\
	\bar{\Phi}_{A\mu} \gamma_5 \widetilde{F}^{\mu\nu} \Phi_{B\nu} &= \frac{1}{2}\bar{\Psi}_{A\mu}\left(F^{\mu\nu} + \gamma_5 \widetilde{F}^{\mu\nu} - \frac{1}{2}\gamma^{\rho\sigma}F_{\rho\sigma}g^{\mu\nu}\right)\Psi_{B\nu}~.
\end{aligned}\end{equation}
Therefore our full action (including gauge-fixing) can be written as
\begin{equation}
	\mathcal{S} = \frac{1}{2\kappa^2}\int d^4x\,\sqrt{-g}\,\bar{\Phi}_{A\mu}\hat{H}^{\mu\nu}_{AB}\Phi_{B\nu}~,
\end{equation}
where
\begin{equation}
	\hat{H}^{\mu\nu}_{AB} =-\gamma^\rho D_\rho g^{\mu\nu}\delta_{AB} + \frac{1}{2}\left(F^{\mu\nu} + \gamma_5 \widetilde{F}^{\mu\nu}\right)\epsilon_{AB}~. 
\end{equation}
Our action is now in the Dirac form required for our heat kernel methods.  We note that the overall normalization in (\ref{eq:gravitino_gauge}) was chosen to enforce this; any other choice would result in an action whose square is non-minimal~\cite{Endo:1994yj}.

As with the hyper fermions we now continue to Euclidean space, giving Hermitian gamma matrices.  The gravitino conjugate is $\bar{\Phi}_{A\mu} = \Phi_{A\mu}^\dagger$, and we will again choose $F^{\mu\nu}$ to be real.  The Hermitian conjugate of $\hat{H}$ is
\begin{equation}
	\hat{H}^{\mu\nu\dagger}_{AB} = \gamma^\rho D_\rho g^{\mu\nu}\delta_{AB} + \frac{1}{2}\left(F^{\mu\nu} - \gamma_5 \widetilde{F}^{\mu\nu}\right)\epsilon_{AB} ~.
\end{equation}
The relevant Laplace-type operator that we will calculate the heat kernel of is
\begin{equation}\begin{aligned}
\label{eq:lambda_gravitino_1}
\Lambda^{\mu\nu}_{AB} &= \hat{H}^{\mu\lambda\dagger}_{AC}\hat{H}_{\lambda~CB}^{~\nu} \\
	&= -\gamma^\rho\gamma^\sigma D_\rho D_\sigma g^{\mu\nu}\delta_{AB} + \frac{1}{4}(F^{\mu\lambda}+\gamma_5 \widetilde{F}^{\mu\lambda})(F_\lambda^{~\nu}-\gamma_5 \widetilde{F}_\lambda^{~\nu})\delta_{AB} \\
	&\quad - \frac{1}{2} \gamma^\rho D_\rho (F^{\mu\nu}-\gamma_5 \widetilde{F}^{\mu\nu})\epsilon_{AB} + \frac{1}{2}(F^{\mu\nu}+\gamma_5 \widetilde{F}^{\mu\nu})\gamma^\rho D_\rho \epsilon_{AB}~.
\end{aligned}\end{equation}
As with the hyper fermions, we can break the two-derivative term $\gamma^\rho\gamma^\sigma D_\rho D_\sigma$ into its symmetric and anti-symmetric parts and use the commutator given in (\ref{eq:commutator_gravitino}).  We will also use the Schouten identity (\ref{eq:schouten}) and gamma matrix commutation relations to simplify this expression.  The result is
\begin{equation}\begin{aligned}
\label{eq:lambda_gravitino}
	\Lambda^{\mu\nu}_{AB} &= -\left( \square g^{\mu\nu}\delta_{AB} + \frac{1}{2}\gamma^{\rho\sigma}R_{\rho\sigma}^{~~\mu\nu}\delta_{AB} + \frac{1}{4}(F^{\mu\lambda}F_\lambda^{~\nu} - \widetilde{F}^{\mu\lambda}\widetilde{F}_\lambda^{~\nu}) \delta_{AB}\right. \\
	&\quad\left. + \frac{1}{2}\gamma^\rho (D_\rho F^{\mu\nu})\epsilon_{AB} - \frac{1}{2}\gamma^\rho \gamma_5 (D_\rho \widetilde{F}^{\mu\nu})\epsilon_{AB} \right)~.
\end{aligned}\end{equation}

In (\ref{eq:lambda_gravitino}) there is no term linear in derivatives.  This corresponds to $\omega_\mu = 0$, and so the matrices $I$ and $E$ are
\begin{equation}
\begin{aligned}
	I^{\mu\nu}_{AB} &= \mathbb{I}_4 g^{\mu\nu}\delta_{AB}~, \\
	E^{\mu\nu}_{AB} &= \left(\frac{1}{2}\gamma^{\rho\sigma}R_{\rho\sigma}^{~~\mu\nu} + \frac{1}{4}F^{\mu\lambda}F_\lambda^{~\nu} - \frac{1}{4}\widetilde{F}^{\mu\lambda}\widetilde{F}_\lambda^{~\nu}\right)\delta_{AB} \\
	&\quad + \left( \frac{1}{2}\gamma^\rho (D_\rho F^{\mu\nu}) - \frac{1}{2}\gamma^\rho (D_\rho \widetilde{F}^{\mu\nu}) \gamma_5 \right)\epsilon_{AB}~.
\end{aligned}
\end{equation}
Since $\omega_\mu = 0$, the curvature $\Omega_{\mu\nu}$ of the connection $\mathcal{D}_\mu$ is given by the commutator in (\ref{eq:commutator_gravitino})
\begin{equation}
	(\Omega_{\rho\sigma})^{\mu\nu}_{AB} = \left( \frac{1}{4}\gamma^{\alpha\beta}R_{\rho\sigma\alpha\beta} g^{\mu\nu} + R_{\rho\sigma}^{~~\mu\nu}\right)\delta_{AB}~.
\end{equation}
The relevant traces for our heat kernel coefficients are
\begin{equation}\begin{aligned}
	\text{Tr }I &= 32~, \\
	\text{Tr }E &= -2F_{\mu\nu}F^{\mu\nu} + 2 \widetilde{F}_{\mu\nu}\widetilde{F}^{\mu\nu}~,  \\
	\text{Tr }E^2 &= 4 R_{\mu\nu\rho\sigma}R^{\mu\nu\rho\sigma} + \frac{1}{2}(F^{\mu\rho}F_{\nu\rho} - \widetilde{F}^{\mu\rho}\widetilde{F}_{\nu\rho})(F_{\mu\sigma}F^{\nu\sigma} - \widetilde{F}_{\mu\sigma}\widetilde{F}^{\nu\sigma}) \\
	&\quad + 2(D^\rho F^{\mu\nu})(D_\rho F_{\mu\nu}) - 2(D^\rho \widetilde{F}^{\mu\nu})(D_\rho \widetilde{F}_{\mu\nu})~, \\
	\text{Tr }\Omega_{\rho\sigma}\Omega^{\rho\sigma} &= -12 R_{\mu\nu\rho\sigma}R^{\mu\nu\rho\sigma}~.
\end{aligned}\end{equation}

We can now calculate the Seeley-DeWitt coefficients (\ref{eq:a2n}) for the gravitini in the gravity multiplet, making sure to add an overall factor of $-1/2$ to account for fermion statistics and the Majorana condition.  We will also simplify the result by using (\ref{eq:dfdf}) to rewrite $(D_\rho F_{\mu\nu})^2$ and $(D_\rho \widetilde{F}_{\mu\nu})^2$ in terms of the Riemann tensor and Ricci tensor contracted with these field strengths.  We end up with
\begin{equation}\begin{aligned}
\label{eq:hk_gravitino}
(4\pi)^2 a_0^{\text{gravitini}}(x) &= -16~, \\
(4\pi)^2 a_2^{\text{gravitini}}(x) &= F_{\mu\nu}F^{\mu\nu} - \widetilde{F}_{\mu\nu}\widetilde{F}^{\mu\nu}~, \\
(4\pi)^2 a_4^{\text{gravitini}}(x) &= -\frac{1}{360}\left(212 R_{\mu\nu\rho\sigma}R^{\mu\nu\rho\sigma} - 32 R_{\mu\nu}R^{\mu\nu} - 360 R_{\mu\nu}(F^{\mu\rho}F^\nu_{~\rho} - \widetilde{F}^{\mu\rho}\widetilde{F}^\nu_{~\rho}) \right. \\
&\quad \left.+ 180 R_{\mu\nu\rho\sigma}(F^{\mu\nu}F^{\rho\sigma}-\widetilde{F}^{\mu\nu}\widetilde{F}^{\rho\sigma})\right. \\
&\quad \left.+45(F^{\mu\rho}F_{\nu\rho} - \widetilde{F}^{\mu\rho}\widetilde{F}_{\nu\rho})(F_{\mu\sigma}F^{\nu\sigma} - \widetilde{F}_{\mu\sigma}\widetilde{F}^{\nu\sigma}) \right)~.
\end{aligned}\end{equation}

As noted in~\cite{Sen:2011ba}, the particular choice of gauge made in (\ref{eq:gravitino_gauge}) induces the ghost Lagrangian
\begin{equation}
\label{eq:gravitino_ghosts}
	\mathcal{L}_\text{ghost} = \bar{\tilde{b}}_A \gamma^\mu D_\mu \tilde{c}_A + \bar{\tilde{e}}_A \gamma^\mu D_\mu \tilde{e}_A~,
\end{equation}
where $\tilde{b}_A$, $\tilde{c}_A$, and $\tilde{e}_A$ are fermionic ghosts, with the same species index $A=1,2$ as the gravitinos.  Since there are six different species of these minimally coupled Majorana fermions, their contribution to the fermionic heat kernel will be $-6$ times the free spin-1/2 heat kernel (\ref{eq:a2n_spinor}).  The net fermionic heat kernel coefficients, including gauge-fixing and ghosts, are $a^{\text{grav},f}_{2n}(x) = a^{\text{gravitini}}_{2n}(x) - 6 a^{1/2}_{2n}(x)$.  The final Seeley-DeWitt coefficients for the fermionic content of the gravity multiplet are thus
\begin{equation}\begin{aligned}
\label{eq:hk_gravity_f}
(4\pi)^2 a_0^{\text{grav},f}(x) &= -4~, \\
(4\pi)^2 a_2^{\text{grav},f}(x) &= F_{\mu\nu}F^{\mu\nu} - \widetilde{F}_{\mu\nu}\widetilde{F}^{\mu\nu}~, \\
(4\pi)^2 a_4^{\text{grav},f}(x) &= -\frac{1}{360}\left(233 R_{\mu\nu\rho\sigma}R^{\mu\nu\rho\sigma} - 8 R_{\mu\nu}R^{\mu\nu} - 360 R_{\mu\nu}(F^{\mu\rho}F^\nu_{~\rho} - \widetilde{F}^{\mu\rho}\widetilde{F}^\nu_{~\rho}) \right. \\
&\quad \left.+ 180 R_{\mu\nu\rho\sigma}(F^{\mu\nu}F^{\rho\sigma}-\widetilde{F}^{\mu\nu}\widetilde{F}^{\rho\sigma})\right. \\
&\quad \left. +45(F^{\mu\rho}F_{\nu\rho} - \widetilde{F}^{\mu\rho}\widetilde{F}_{\nu\rho})(F_{\mu\sigma}F^{\nu\sigma} - \widetilde{F}_{\mu\sigma}\widetilde{F}^{\nu\sigma}) \right)~.
\end{aligned}\end{equation}

%%%%%%%%%%%%%%%%%%%%%%%%%%%%%%%%%%%%%%%%%%%%%%%%%%%%%%%%%%%%%%

\subsubsection{Gravity Multiplet: Bosons}
\label{subsubsec:heatkernel_n=2_gravitymultiplet_b}

As discussed in~\S\ref{subsec:background_n=2sugra}, the action for the bosonic content of the gravity multiplet coincides with the Einstein-Maxwell action
\begin{equation}
\label{eq:gravbaction}
	\mathcal{S} = \frac{1}{2\kappa^2}\int d^4x\,\sqrt{-g}\,\left(R - \frac{1}{4}F_{\mu\nu}F^{\mu\nu}\right)~,
\end{equation}
where $R$ is the Ricci scalar corresponding to the metric $g_{\mu\nu}$ and $F_{\mu\nu} = D_\mu A_\nu - D_\nu A_\mu$ is the background graviphoton field strength.  We want to consider quadratic fluctuations about the background and then compute the corresponding heat kernel.  This calculation has been done for Einstein-Maxwell theory~\cite{Bhattacharyya:2012wz}, but we find it useful to go through it in detail.

Consider the variations
\begin{equation}
	\delta g_{\mu\nu} = h_{\mu\nu}~,\quad \delta A_\mu = a_\mu~.
\end{equation}
The fluctuations are the graviton $h_{\mu\nu}$ and the graviphoton $a_\mu$. We will expand the action (\ref{eq:gravbaction}) to quadratic order in these field fluctuations.  The relevant second-order variations, up to a total derivative, are
\begin{equation}\begin{aligned}
\label{eq:variations}
	\delta^2 \left( \sqrt{-g}R \right) &= \sqrt{-g} \bigg{[} \frac{1}{2}h^{\mu\nu}\square h_{\mu\nu} - \frac{1}{2}h^\mu_{~\mu} \square h^\rho_{~\rho} + h^{\mu\nu}h^{\rho\sigma}R_{\mu\rho\nu\sigma} + h^{\mu\nu}h_{\mu\rho}R^\rho_{~\nu} \\
	&\quad\quad\quad\quad + \frac{1}{4} (h^\mu_{~\mu})^2R - h^{\mu\nu}h^\rho_{~\rho}R_{\mu\nu} - \frac{1}{2}h^{\mu\nu}h_{\mu\nu}R  \\
	&\quad\quad\quad\quad +(D^\mu h_{\mu\nu})(D^\rho h_{\rho}^{~\nu}) + (D^\mu D^\nu h_{\mu\nu})h^\rho_{~\rho} \bigg{]}~,\\
	\delta^2 \left( \sqrt{-g}F_{\mu\nu}F^{\mu\nu} \right) &= \sqrt{-g} \bigg{[} 2 f_{\mu\nu}f^{\mu\nu} - \frac{1}{2}\left(h^{\mu\nu}h_{\mu\nu} - \frac{1}{2}(h^\mu_{~\mu})^2\right) F_{\rho\sigma} F^{\rho\sigma}  \\
	&\quad\quad\quad\quad -8 h^{\mu\nu}f_{\mu\rho}F_\nu^{~\rho} + 2 h^\rho_{~\rho} f_{\mu\nu}F^{\mu\nu} +4 h^{\mu\nu}h^\rho_{~\nu}F_{\mu\sigma}F_\rho^{~\sigma}  \\
	&\quad\quad\quad\quad + 2 h^{\mu\nu}h^{\rho\sigma}F_{\mu\rho}F_{\nu\sigma} - 2 h^{\mu\nu}h^\rho_{~\rho}F_{\mu\sigma}F_\nu^{~\sigma}\bigg{]}~,
\end{aligned}\end{equation}
where $f_{\mu\nu} = D_\mu a_\nu - D_\nu a_\mu$.  We gauge-fix our theory by
\begin{equation}
	\mathcal{L}_{\text{g.f.}} = -\frac{1}{2\kappa^2}\left(D^\mu h_{\mu\rho} - \frac{1}{2} D_\rho h^\mu_{~\mu}\right)\left(D^\nu h_{\nu\sigma} - \frac{1}{2} D_\sigma h^\nu_{~\nu}\right) -  \frac{1}{2\kappa^2}(D^\mu a_\mu)^2~,
\label{eq:gravgf}
\end{equation}
which picks out the harmonic gauge for the graviton ($D^\mu h_{\mu\rho} - \frac{1}{2} D_\rho h^\mu_{~\mu} = 0$) and the Lorenz gauge for the graviphoton ($D^\mu a_\mu = 0$).  We use the background Einstein equations to simplify the gauge-fixed quadratic action, which includes setting $R=0$.  Additionally, we let $h_{\mu\nu}\rightarrow \sqrt{2}h_{\mu\nu}$ so that the kinetic terms for the graviton and the graviphoton have the same normalization.  The resulting action is
\begin{equation}\begin{aligned}
	\mathcal{S} &= \frac{1}{2\kappa^2}\int d^4x\,\sqrt{-g}\,\bigg{[}h^{\mu\nu}\square h_{\mu\nu} - \frac{1}{2}h^\mu_{~\mu} \square h^\rho_{~\rho} + a^\mu\left(\square g_{\mu\nu} - R_{\mu\nu}\right)a^\nu + 2h^{\mu\nu}h^{\rho\sigma}R_{\mu\rho\nu\sigma}\\
	&\quad  - 2h^{\mu\nu}h_{\mu\rho}R^\rho_{~\nu} - \frac{1}{4}h^{\mu\nu}h_{\mu\nu}F_{\rho\sigma}F^{\rho\sigma} + \frac{1}{8}(h^\mu_{~\mu})^2F_{\rho\sigma}F^{\rho\sigma} - h^{\mu\nu}h^{\rho\sigma}F_{\mu\rho}F_{\nu\sigma} \\
	&\quad - \frac{1}{\sqrt{2}}h^\rho_{~\rho}f_{\mu\nu}F^{\mu\nu} + 2\sqrt{2} h^{\mu\nu}f_{\mu\rho}F_\nu^{~\rho}\bigg{]}~.
\label{eq:quadgrav1}
\end{aligned}\end{equation}

We note that (\ref{eq:quadgrav1}) is not in the required Laplace form needed for our heat kernel analysis, due to the $h^\mu_{~\mu}\square h^\rho_{~\rho}$ kinetic term.  To fix this, we separate the graviton $h_{\mu\nu}$ into its trace $h$ and traceless component $\phi_{\mu\nu}$ by defining
\begin{equation}
	h \equiv h^\mu_{~\mu}~,
\end{equation}
\begin{equation}
	\phi_{\mu\nu} \equiv h_{\mu\nu} - \frac{1}{4}g_{\mu\nu} h~.
\end{equation}
This decomposition is standard, as the fields $h$ and $\phi_{\mu\nu}$ transform under irreducible representations of $SL(2,\mathbb{C})$~\cite{Christensen:1979iy,Christensen:1978md,Christensen:1978gi,Gibbons:1978ji}.  The action becomes
\begin{equation}\begin{aligned}
	\mathcal{S} &= \frac{1}{2\kappa^2}\int d^4x\,\sqrt{-g}\,\bigg{[}\phi^{\mu\nu}\square \phi_{\mu\nu} - \frac{1}{4}h \square h + a^\mu\left(\square g_{\mu\nu} - R_{\mu\nu}\right)a^\nu + 2\phi^{\mu\nu}\phi^{\rho\sigma}R_{\mu\rho\nu\sigma}\\
	&\quad  - 2\phi^{\mu\nu}\phi_{\mu\rho}R^\rho_{~\nu} - \frac{1}{4}\phi^{\mu\nu}\phi_{\mu\nu}F_{\rho\sigma}F^{\rho\sigma} - \phi^{\mu\nu}\phi^{\rho\sigma}F_{\mu\rho}F_{\nu\sigma} -h \phi_{\mu\nu}R^{\mu\nu} + 2\sqrt{2} \phi^{\mu\nu}f_{\mu\rho}F_\nu^{~\rho}\bigg{]}~.
\end{aligned}\end{equation}

The kinetic term for $h$ has a negative sign.  This is the conformal factor problem in gravity, and results in an unbounded path integral for our theory.  The resolution to this problem is that the one-loop effective action can be made to converge by performing a conformal rotation that takes the contour of integration for $h$ to be along the imaginary axis~\cite{Gibbons:1978ac,Mazur:1989by,Schleich:1987fm}.  We will also simultaneously rescale $h$ to make the normalization of its kinetic term coincide with those for $\phi_{\mu\nu}$ and $a_\mu$.  Therefore, we let
\begin{equation}
	\phi = -\frac{i}{2} h~,
\end{equation} 
and consider the action quadratic in the fields $\{\phi_n\} = \{\phi_{\mu\nu}, a_\mu, \phi\}$. The result is
\begin{equation}
\mathcal{S} = -\frac{1}{2\kappa^2}\int d^4x\,\sqrt{-g}\, \phi_n \Lambda^n_m \phi^m ~,
\end{equation}
where $\Lambda$ acts on our fields by
\begin{equation}\begin{aligned}
\label{eq:quadgrav2}
	-\phi_n \Lambda^n_m \phi^m &= \phi_{\mu\nu}\left( \square g^\mu_{~\rho}g^\nu_{~\sigma} - 2 R^{\mu}_{~\rho}g^\nu_{~\sigma} + 2R^{\mu~\nu}_{~\rho~\sigma} - \frac{1}{4}g^\mu_{~\rho}g^\nu_{~\sigma} F_{\lambda\tau}F^{\lambda\tau} - F^\mu_{~\rho}F^\nu_{~\sigma}\right)\phi^{\rho\sigma} \\
	&\quad + a_\mu\left(\square g^{\mu}_{~\rho} - R^\mu_{~\rho}\right)a^\rho + \phi \square \phi + \phi_{\mu\nu}\left(-i R^{\mu\nu}\right)\phi + \phi \left(-i R_{\rho\sigma}\right)\phi^{\rho\sigma} \\
	&\quad + \phi_{\mu\nu}\left( \frac{\sqrt{2}}{2}(D^\mu F_\rho^{~\nu}) + \sqrt{2} (F_\alpha^{~\nu} g^\mu_{~\rho}-F_\rho^{~\nu}g^\mu_{~\alpha})D^\alpha \right) a^\rho \\
	&\quad + a_\mu\left( \frac{\sqrt{2}}{2}(D_\rho F^\mu_{~\sigma}) + \sqrt{2}(F^\mu_{~\sigma}g_{\rho\alpha} - F_{\alpha\sigma}g_\rho^{~\mu})D^\alpha \right)\phi^{\rho\sigma}~.
\end{aligned}\end{equation}
We have adjusted total derivative terms to make $\Lambda$ Hermitian.  From (\ref{eq:quadgrav2}), the matrices $P$ and $\omega_\alpha$ are
\begin{equation}\begin{aligned}
	\phi_n P^n_m \phi^m &= \phi_{\mu\nu}\left(- 2 R^{\mu}_{~\rho}g^\nu_{~\sigma} + 2R^{\mu~\nu}_{~\rho~\sigma} - \frac{1}{4}g^\mu_{~\rho}g^\nu_{~\sigma} F_{\lambda\tau}F^{\lambda\tau} - F^\mu_{~\rho}F^\nu_{~\sigma}  \right)\phi^{\rho\sigma} \\
	&\quad + a_\mu\left(-R^\mu_{~\rho}\right)a^\rho + \phi_{\mu\nu}\left(-i R^{\mu\nu}\right)\phi + \phi\left(-i R_{\rho\sigma}\right)\phi^{\rho\sigma} \\
	&\quad + \phi_{\mu\nu}\left(\frac{\sqrt{2}}{2}(D^\mu F_\rho^{~\nu})\right) a^\rho + a_\mu\left( \frac{\sqrt{2}}{2}(D_\rho F^\mu_{~\sigma}) \right)\phi^{\rho\sigma}~,
\end{aligned}\end{equation}
\begin{equation}\begin{aligned}
	\phi_n(\omega_\alpha)^n_m \phi^m &= \frac{\sqrt{2}}{2} \phi_{\mu\nu}\left( F_\alpha^{~\nu}g^\mu_{~\rho} - F_{\rho}^{~\nu}g^\mu_{~\alpha} \right)a^\rho + \frac{\sqrt{2}}{2} a_\mu\left(F^\mu_{~\sigma}g_{\rho\alpha} - F_{\alpha\sigma}g_\rho^{~\mu} \right)\phi^{\rho\sigma}~.
\end{aligned}\end{equation}

We now define the operator
\begin{equation}
\label{eq:tracelessproject}
	G^{\mu\nu}_{\rho\sigma} = \frac{1}{2}\left(g^\mu_{~\rho}g^\nu_{~\sigma} + g^\mu_{~\sigma}g^\nu_{~\rho} - \frac{1}{2}g^{\mu\nu}g_{\rho\sigma}\right)~.
\end{equation}
$G^{\mu\nu}_{\rho\sigma}$ projects onto the traceless part of a symmetric tensor.  In order to impose that $\phi_{\mu\nu}$ is the traceless part of the graviton, we must use $G^{\mu\nu}_{\rho\sigma}$ to contract pairs of indices for any operator acting on $\phi_{\mu\nu}$.  That is, if we have some matrix $M$ acting on our fields such that
\begin{equation}
	\phi_n M^n_m \phi^m = \phi_{\mu\nu}M^{\mu\nu}_{\rho\sigma}\phi^{\rho\sigma}~,
\end{equation}
then $M^2$ is given by
\begin{equation}
	\phi_n (M^2)^n_m \phi^m = \phi_{\mu\nu}M^{\mu\nu}_{\alpha\beta}G^{\alpha\beta}_{\gamma\delta}M^{\gamma\delta}_{\rho\sigma}\phi^{\rho\sigma}~.
\end{equation}
We must also use $G^{\mu\nu}_{\rho\sigma}$ when taking traces of these operators, i.e.
\begin{equation}
	\text{Tr }M = G^{\rho\sigma}_{\mu\nu}M^{\mu\nu}_{\rho\sigma}~.
\end{equation}
As an example, the identity operator $I_g$ acting on $\phi_{\mu\nu}$ is defined by
\begin{equation}
	\phi_n (I_g)^n_m \phi^m = \phi_{\mu\nu}\left(g^\mu_{~\rho}g^\nu_{~\sigma}\right)\phi^{\rho\sigma} = \phi_{\mu\nu}\phi^{\mu\nu}~.
\end{equation}
Since $\phi_{\mu\nu}$ is both symmetric and traceless, we expect it to have $10-1=9$ independent off-shell degrees of freedom. The trace of $I_g$, using $G^{\mu\nu}_{\rho\sigma}$ to contract indices, is indeed
\begin{equation}
	\text{Tr }I_g = G^{\rho\sigma}_{\mu\nu}g^\mu_{~\rho}g^\nu_{~\sigma} = G^{\mu\nu}_{\mu\nu} = \frac{1}{2}\left( g^\mu_{~\mu}g^\nu_{~\nu} + \frac{1}{2}g^{\mu\nu}g_{\mu\nu}\right) = 9~.
\end{equation}

Using the traceless projection operator (\ref{eq:tracelessproject}) with our expressions for $P$ and $\omega_\alpha$ and the background equations of motion, it follows that $\omega^\alpha\omega_\alpha$ and $(D^\alpha\omega_\alpha)$ are
\begin{equation}\begin{aligned}
	\phi_n (\omega^\alpha \omega_\alpha)^n_m \phi^m &= \phi_{\mu\nu}\left( -F^\mu_{~\rho}F^\nu_{~\sigma}-2R^\mu_{~\rho}g^\nu_{~\sigma} - \frac{1}{4}g^\mu_{~\rho}g^\nu_{~\sigma}F_{\lambda\tau}F^{\lambda\tau} \right)\phi^{\rho\sigma} \\
	&\quad + a_\mu\left(-R^\mu_{~\rho}-\frac{3}{8}g^{\mu}_{~\rho}F_{\lambda\tau}F^{\lambda\tau}\right)a^\rho~,
\end{aligned}\end{equation}
\begin{equation}\begin{aligned}
	\phi_n (D^\alpha \omega_{\alpha})^n_m \phi^m &= \phi_{\mu\nu}\left( -\frac{\sqrt{2}}{2}(D^\mu F_\rho^{~\nu}) \right) a^\rho + a_\mu\left( \frac{\sqrt{2}}{2}(D_\rho F^\mu_{~\sigma}) \right)\phi^{\rho\sigma}~.
\end{aligned}\end{equation}
Using $E = P - \omega^\alpha\omega_\alpha - (D^\alpha\omega_\alpha)$ and adjusting total derivative terms to make $E$ Hermitian, we find that
\begin{equation}\label{eq:grave}
\begin{aligned}
	\phi_n E^n_m \phi^m &= \phi_{\mu\nu}\left( 2R^{\mu~\nu}_{~\rho~\sigma} \right)\phi^{\rho\sigma}  + a_\mu\left(\frac{3}{8}g^\mu_{~\rho}F_{\lambda\tau}F^{\lambda\tau}\right)a^\rho \\
	&\quad + \phi_{\mu\nu}(-iR^{\mu\nu})\phi + \phi(-iR_{\rho\sigma})\phi^{\rho\sigma} \\
	&\quad + \phi_{\mu\nu}\left( \frac{\sqrt{2}}{2}(D^\mu F_\rho^{~\nu}) \right)a^\rho + a_\mu \left( \frac{\sqrt{2}}{2}(D_\rho F^\mu_{~\sigma}) \right) \phi^{\rho\sigma}~.
\end{aligned}\end{equation}

The traceless graviton $\phi_{\mu\nu}$ has nine off-shell degrees of freedom, while the trace $\phi$ has only one and the graviphoton $a_\mu$ has four.  Therefore,
\begin{equation}
\text{Tr }I = 9 + 1 + 4 = 14~.
\end{equation}
From (\ref{eq:grave}) it follows that
\begin{equation}\begin{aligned}
\label{eq:grav_tre}
\text{Tr }E &= \frac{3}{2}F_{\mu\nu}F^{\mu\nu}~, \\
\text{Tr }E^2 &= 3 R_{\mu\nu\rho\sigma}R^{\mu\nu\rho\sigma} - 7 R_{\mu\nu}R^{\mu\nu} + \frac{3}{4}R_{\mu\nu\rho\sigma}F^{\mu\nu}F^{\rho\sigma} + \frac{9}{16}(F_{\mu\nu}F^{\mu\nu})^2~.
\end{aligned}\end{equation}
In order to compute the curvature $\Omega_{\alpha\beta}$ we expand the commutator in (\ref{eq:omeg}):
\begin{equation}
	(\Omega_{\alpha\beta})^n_m = [D_\alpha,D_\beta]^n_m + \phi_n( D_{[\alpha}\omega_{\beta]})^n_m\phi^m + [\omega_\alpha,\omega_\beta]^n_m~.
\label{eq:om3}
\end{equation}
The covariant derivative commutes when acting on $\phi$ but not when acting on $a_\mu$ or $\phi_{\mu\nu}$.  We also account for the fact that $\phi_{\mu\nu}$ is symmetric.  So, the first term in (\ref{eq:om3}) is
\begin{equation}
\begin{aligned}
	\phi_n[D_\alpha,D_\beta]^n_m \phi^m &= \phi_{\mu\nu}[D_\alpha,D_\beta]\phi^{\mu\nu} + a_\mu [D_\alpha,D_\beta]a^\mu \\
	&= \phi_{\mu\nu}\left(2 R^\mu_{~\rho\alpha\beta}g^\nu_{~\sigma}\right)\phi^{\rho\sigma} + a_\mu(R^\mu_{~\rho\alpha\beta})a^\rho~.
\end{aligned}
\end{equation}
The second term in (\ref{eq:om3}) can be calculated by applying the covariant derivative to $\omega_\alpha$ and simplifying with the Bianchi identity:
\begin{equation}\begin{aligned}
	\phi_n( D_{[\alpha}\omega_{\beta]})^n_m\phi^m &= \frac{\sqrt{2}}{2}\phi_{\mu\nu}\left( -D^\nu F_{\alpha\beta}g^\mu_{~\rho} - g^\mu_{~[\beta}D_{\alpha]}F_\rho^{~\nu}\right) a^\rho \\
	&\quad+ \frac{\sqrt{2}}{2}a_\mu\left( D_\sigma F_{\alpha\beta}g^\mu_{~\rho} +g_{\rho[\beta}D_{\alpha]}F^\mu_{~\sigma} \right)\phi^{\rho\sigma}~.
\end{aligned}\end{equation}
Note that the covariant derivative is applied only to the background field strength tensors in the above expression, and not to the fields themselves.  The last term in equation (\ref{eq:om3}) is obtained by taking a product of $\omega_\alpha$ and $\omega_\beta$, antisymmetrizing, and simplifying with the background equations of motion, giving
\begin{equation}\begin{aligned}
	\phi_n [\omega_\alpha,\omega_\beta]^n_m\phi^m &= \frac{1}{2}\phi_{\mu\nu}\big{(}F^\mu_{~\rho}F_\alpha^{~\nu}g_{\beta\sigma} - F^\mu_{~\rho}F_{\beta\sigma}g_\alpha^{~\nu} - F_\alpha^{~\nu}F_{\beta\sigma}g^\mu_{~\rho} - 2 R^\mu_{~\rho}g_\alpha^{~\nu}g_{\beta\sigma} \\
	&\quad - \frac{1}{4}g^\mu_{~\rho}g_\alpha^{~\nu}g_{\beta\sigma}F_{\lambda\tau}F^{\lambda\tau}\big{)}\phi^{\rho\sigma} + \frac{1}{2} a_\mu\left(R_\beta^{~\mu}g_{\alpha\rho} + R_{\alpha\rho}g_\beta^{~\mu} -F^\mu_{~\rho}F_{\alpha\beta} \right)a^\rho \\
	&\quad-(\alpha \leftrightarrow \beta)~.
\end{aligned}\end{equation}
We have all of the components of $\Omega_{\mu\nu}$ and so it is straightforward to compute the trace of $\Omega_{\mu\nu}\Omega^{\mu\nu}$ using the background equations of motion and Bianchi identities.  The result, up to a total derivative, is
\begin{equation}
\label{eq:grav_tro2}
\text{Tr }\Omega_{\mu\nu}\Omega^{\mu\nu} = -7 R_{\mu\nu\rho\sigma}R^{\mu\nu\rho\sigma} + 56 R_{\mu\nu}R^{\mu\nu} - \frac{9}{2}R_{\mu\nu\rho\sigma}F^{\mu\nu}F^{\rho\sigma}-\frac{27}{8}(F_{\mu\nu}F^{\mu\nu})^2~.
\end{equation}

The choice of gauge-fixing in (\ref{eq:gravgf}) induces the ghost Lagrangian
\begin{equation}
\label{eq:lghost1}
	\mathcal{L}_\text{ghost} = 2b_\mu (\square g^{\mu\nu} + R^{\mu\nu})c_\nu + 2 b\square c -4 b F^{\mu\nu}D_\mu c_\nu~,
\end{equation}
where $b_\mu,c_\mu$ are the diffeomorphism ghosts associated with the graviton and $b,c$ are the ghosts associated with the graviphoton.  For the purposes of computing the heat kernel coefficients we can treat $b_\mu,c_\mu$ as vector fields and $b,c$ as scalar fields.  In order to make the kinetic term in (\ref{eq:lghost1}) diagonal, we make the change of variables
\begin{equation}
	b'_\mu = \frac{i(b_\mu - c_\mu)}{\sqrt{2}}~,\quad c'_\mu = \frac{b_\mu+c_\mu}{\sqrt{2}}~,\quad b' = \frac{i(b-c)}{\sqrt{2}}~,\quad c'=\frac{b+c}{\sqrt{2}}~.
\end{equation}
If we insert these into (\ref{eq:lghost1}) and adjust the total-derivative terms to make the action Hermitian, we find that
\begin{equation}\begin{aligned}
	S &= \int d^4x\,\sqrt{-g}\,\bigg{[} c'_\mu(\square g^{\mu\nu} + R^{\mu\nu})c'_\nu + b'_\mu(\square g^{\mu\nu} + R^{\mu\nu})b'_\nu + b'\square b' + c\square c' \\
	&\quad -  (b'_\mu - i c'_\mu)F^{\mu\nu}D_\nu(b+ic) -(b+ic)F^{\mu\nu}D_\mu(b'_\nu - i c'_\nu)\bigg{]}~.
\end{aligned}\end{equation}
We will now supress the $'$ on these terms for notational simplicity.  From this action, we can read off the matrices $P$ and $\omega_\alpha$ as
\begin{equation}\begin{aligned}
\label{eq:useghost1}
	\phi_n P^n_m \phi^m &= b_\mu(R^\mu_{~\nu})b^\nu + c_\mu(R^\mu_{~\nu})c^\nu~, \\
	\phi_n (\omega_\alpha)^n_m \phi^m &= -\frac{1}{2} (b_\mu - i c_\mu)F^\mu_{~\alpha}(b+ic) -\frac{1}{2} (b+ic)F_{\alpha\nu}(b^\nu - i c^\nu)~.
\end{aligned}\end{equation}
The commutator of two covariant derivatives commutes when acting on the scalar ghosts but not on the vector ghosts, so
\begin{equation}
\label{eq:useghost2}
	\phi_n [D_\alpha,D_\beta]^n_m \phi^m = b_\mu(R^\mu_{~\nu\alpha\beta})b^\nu + c_\mu(R^\mu_{~\nu\alpha\beta})c^\nu~.
\end{equation}
Using (\ref{eq:useghost1}) and (\ref{eq:useghost2}) it is straightforward to compute $E$ and $\Omega_{\alpha\beta}$ for the ghosts:
\begin{equation}\begin{aligned}
	\phi_n E^n_m \phi^m &= b_\mu(R^\mu_{~\nu})b^\nu + c_\mu(R^\mu_{~\nu})c^\nu~, \\
	\phi_n (\Omega_{\alpha\beta})^n_m \phi^m &= b_\mu(R^\mu_{~\nu\alpha\beta})b^\nu + c_\mu(R^\mu_{~\nu\alpha\beta})c^\nu  - \frac{1}{2}(b_\mu - i c_\mu)(D^\mu F_{\alpha\beta})(b+ic)\\
	&\quad +\frac{1}{2} (b+ic)(D_\nu F_{\alpha\beta})(b^\nu - i c^\nu)~.
\end{aligned}\end{equation}
Each vector ghost has four degrees of freedom, while the scalars each have one, giving $\text{Tr }I = 4+4+1+1 = 10$. The traces of $E$, $E^2$, and $\Omega_{\mu\nu}\Omega^{\mu\nu}$ are
\begin{equation}
\label{eq:ghosts_trace}
\text{Tr }E = 0~, \quad \text{Tr }E^2 = 2 R_{\mu\nu}R^{\mu\nu}~, \quad \text{Tr }\Omega_{\mu\nu}\Omega^{\mu\nu} = -2 R_{\mu\nu\rho\sigma}R^{\mu\nu\rho\sigma}~.
\end{equation}

The total Seeley-DeWitt coefficients for the bosons in the gravity multiplet are given by inserting the traces in (\ref{eq:grav_tre}) and (\ref{eq:grav_tro2}) (as well as the ghost traces in (\ref{eq:ghosts_trace}) with an overall minus sign) into (\ref{eq:a2n}).  The result is
\begin{equation}
\label{eq:hk_gravity_b}
\begin{aligned}
(4\pi)^2 a_0^{\text{grav},b}(x) &= 4~, \\
(4\pi)^2 a_2^{\text{grav},b}(x) &= \frac{3}{2}F_{\mu\nu}F^{\mu\nu}~, \\
(4\pi)^2 a_4^{\text{grav},b}(x) &= \frac{1}{180}\left( 199 R_{\mu\nu\rho\sigma}R^{\mu\nu\rho\sigma} + 26 R_{\mu\nu}R^{\mu\nu}\right)~.
\end{aligned}
\end{equation}
$a_4^{\text{grav},b}(x)$ matches exactly with the Einstein-Maxwell result given in~\cite{Bhattacharyya:2012wz}.  As mentioned there, it has no explicit dependence on the background graviphoton field strength, although we would have obtained a different result if we had ignored the terms involving the field strength in the action.

The full gravity multiplet heat kernel coefficients, with contributions from the graviton, gravitini, and graviphoton fluctuations, are
\begin{equation}
\label{eq:hk_gravity}
\begin{aligned}
(4\pi)^2 a_0^{\text{grav}}(x) &= 0~, \\
(4\pi)^2 a_2^{\text{grav}}(x) &= \frac{5}{2}F_{\mu\nu}F^{\mu\nu} - \widetilde{F}_{\mu\nu}\widetilde{F}^{\mu\nu}~, \\
(4\pi)^2 a_4^{\text{grav}}(x) &= \frac{1}{24}\left(11 R_{\mu\nu\rho\sigma}R^{\mu\nu\rho\sigma} + 4 R_{\mu\nu}R^{\mu\nu} +24 R_{\mu\nu}(F^{\mu\rho}F^\nu_{~\rho} - \widetilde{F}^{\mu\rho}\widetilde{F}^\nu_{~\rho}) \right. \\
&\quad \left.-12 R_{\mu\nu\rho\sigma}(F^{\mu\nu}F^{\rho\sigma}-\widetilde{F}^{\mu\nu}\widetilde{F}^{\rho\sigma})\right. \\
&\quad \left. -3(F^{\mu\rho}F_{\nu\rho} - \widetilde{F}^{\mu\rho}\widetilde{F}_{\nu\rho})(F_{\mu\sigma}F^{\nu\sigma} - \widetilde{F}_{\mu\sigma}\widetilde{F}^{\nu\sigma}) \right)~.
\end{aligned}\end{equation}

%%%%%%%%%%%%%%%%%%%%%%%%%%%%%%%%%%%%%%%%%%%%%%%%%%%%%%%%%%%%%%

\subsubsection{Gravitino Multiplet}
\label{subsubsec:heatkernel_n=2_gravitinomultiplet}

By gravitino multiplet we refer to the additional $\mathcal{N}-2$ gravitini, referred to as massive gravitini, and their superpartners in $\mathcal{N}=2$ SUSY.  The $\mathcal{N}=2$ gravitino multiplet consists of a (massive) Majorana gravitino, two vector fields, and a Majorana gaugino.  The two vector fields are minimally coupled to gravity, so the heat kernel coefficients (including ghosts resulting from the standard Lorenz gauge-fixing) are well-known~\cite{Vassilevich:2003xt,Birrell:1982}:
\begin{equation}\begin{aligned}
\label{eq:a2n_3/2_b}
	(4\pi)^2a_0^{3/2,b}(x) &= 4~, \\
	(4\pi)^2a_2^{3/2,b}(x) &= 0~, \\
	(4\pi)^2a_4^{3/2,b}(x) &= \frac{1}{90}\left(-13 R_{\mu\nu\rho\sigma}R^{\mu\nu\rho\sigma} + 88 R_{\mu\nu}R^{\mu\nu}\right)~. \\
\end{aligned}\end{equation} 

The fermions in the gravitino multiplet are coupled together by the background graviphoton field.  The Lagrangian describing these interactions is given in (\ref{eq:ee}) as
\begin{equation}
	\mathcal{L}_\text{gravitino} = -\frac{1}{\kappa^2}\bar{\Psi}_\mu\gamma^{\mu\nu\rho}D_\nu \Psi_\rho - \frac{2}{\kappa^2}\bar{\lambda}\gamma^\mu D_\mu \lambda - \frac{1}{2\kappa^2}\left(\bar{\Psi}_\mu\hat{F}\gamma^\mu \lambda + \bar{\lambda}\gamma^\mu \hat{F}\Psi_\mu\right)~,
\end{equation}
where $\Psi_\mu$ is the gravitino field, $\lambda$ is the gaugino field, and $\hat{F} = \frac{1}{2}F_{\mu\nu}\gamma^{\mu\nu}$.  We will proceed as we did for the gravitini in the gravity multiplet and choose the harmonic gauge $\gamma^\mu \Psi_\mu = 0$ by adding to our Lagrangian the gauge-fixing term
\begin{equation}
	\mathcal{L}_\text{g.f.} = \frac{1}{2\kappa^2}(\bar{\Psi}_\mu\gamma^\mu)\gamma^\nu D_\nu(\gamma^\rho \Psi_\rho)~.
\end{equation}
We also make the field redefinition
\begin{equation}
	\Phi_\mu = \frac{1}{\sqrt{2}}\Psi_\mu - \frac{1}{2\sqrt{2}}\gamma_\mu \gamma^\nu \Psi_\nu~.
\end{equation}
Let $\{\phi_m\} = \{\Phi_\mu,\lambda\}$ denote our fermionic fields.  Then, the total action quadratic in these fields (including gauge-fixing) is
\begin{equation}
\label{eq:mgravaction}
	\mathcal{S} = \frac{1}{\kappa^2}\int d^4x\,\sqrt{-g}\,\phi_n \hat{H}^n_m \phi^m~,
\end{equation}
where
\begin{equation}
	\phi_n \hat{H}^n_m \phi^m = -\bar{\Phi}_\mu \gamma^\nu D_\nu \Phi^\mu - \bar{\lambda}\gamma^\nu D_\nu \lambda - \frac{\sqrt{2}}{4}\left( \bar{\Phi}_\mu \hat{F}\gamma^\mu\lambda + \bar{\lambda}\gamma^\mu\hat{F}\Phi_\mu\right)~.
\end{equation}
The action (\ref{eq:mgravaction}) is in the Dirac form needed to employ our heat kernel methods, where $\hat{H}$ is the Dirac-type operator acting on our fermionic fields.

From here, the story is familiar: we continue to Euclidean space, take $\Lambda = \hat{H}\hat{H}^\dagger$, and compute the heat kernel of $\Lambda$ using all of our standard tricks.  We will also include the ghost contribution (see (\ref{eq:gravitino_ghosts})) that results from our choice of gauge-fixing and subtract that from the massive gravitino and gaugino contribution.  The resulting heat kernel coefficients are
\begin{equation}\begin{aligned}
\label{eq:a2n_3/2_f}
	(4\pi^2)a_0^{3/2,f}(x) &= -4~, \\
	(4\pi^2)a_2^{3/2,f}(x) &= -F_{\mu\nu}F^{\mu\nu}~, \\
	(4\pi^2)a_4^{3/2,f}(x) &= -\frac{1}{360}\bigg{[}113R_{\mu\nu\rho\sigma}R^{\mu\nu\rho\sigma} - 8 R_{\mu\nu}R^{\mu\nu} - 15R_{\mu\nu\rho\sigma}\left(F^{\mu\nu}F^{\rho\sigma}-\widetilde{F}^{\mu\nu}\widetilde{F}^{\rho\sigma}\right) \\
	&\quad\quad\quad	- \frac{45}{4}\left((F_{\mu\nu}F^{\mu\nu})^2 - (F_{\mu\nu}\widetilde{F}^{\mu\nu})^2 \right)\bigg{]}~.
\end{aligned}\end{equation}
Adding up the bosonic (\ref{eq:a2n_3/2_b}) and fermionic (\ref{eq:a2n_3/2_f}) contributions, the net heat kernel coefficients for the massive gravitino multiplet are
\begin{equation}\begin{aligned}
\label{eq:a2n_3/2}
	(4\pi^2)a_0^{3/2}(x) &= 0~, \\
	(4\pi^2)a_2^{3/2}(x) &= -F_{\mu\nu}F^{\mu\nu}~, \\
	(4\pi^2)a_4^{3/2}(x) &=\frac{1}{24}\bigg{[}-11R_{\mu\nu\rho\sigma}R^{\mu\nu\rho\sigma} + 24 R_{\mu\nu}R^{\mu\nu} + R_{\mu\nu\rho\sigma}\left(F^{\mu\nu}F^{\rho\sigma}-\widetilde{F}^{\mu\nu}\widetilde{F}^{\rho\sigma}\right) \\
	&\quad\quad\quad	+ \frac{3}{4}\left((F_{\mu\nu}F^{\mu\nu})^2 - (F_{\mu\nu}\widetilde{F}^{\mu\nu})^2 \right)\bigg{]}~.
\end{aligned}\end{equation}

%%%%%%%%%%%%%%%%%%%%%%%%%%%%%%%%%%%%%%%%%%%%%%%%%%%%%%%%%%%%%%

\section{Discussion}
\label{sec:discussion}

In this section we collect our results and simplify their form.  We compute the corresponding logarithmic corrections to black hole entropy.  We discuss the significance of our results and the implications for Kerr-Newman black holes.

%%%%%%%%%%%%%%%%%%%%%%%%%%%%%%%%%%%%%%%%%%%%%%%%%%%%%%%%%%%%%%

\subsection{The Conformal Anomaly and Central Charges}
\label{subsec:discussion_conformal}

The $a_4(x)$ coefficients derived in each $\mathcal{N}=2$ multiplet are linear combinations of covariant terms that each contain four derivatives
\begin{equation}
	a_4(x) = \alpha_1 R_{\mu\nu\rho\sigma}R^{\mu\nu\rho\sigma} + \alpha_2 R_{\mu\nu}R^{\mu\nu} + \alpha_3 R_{\mu\nu\rho\sigma}F^{\mu\nu}F^{\rho\sigma} + ...~,
\end{equation}
for some set of numerical coefficients $\{\alpha_i\}$.  We found it useful to keep $F_{\mu\nu}$ and $\widetilde{F}_{\mu\nu}$ distinct in~\S\ref{sec:heatkernel} but we will now simplify as much as possible by expressing the dual field strength in terms of the Levi-Civita symbol and the field strength.  We use the Einstein equation (\ref{eq:einstein_r=0}), the Schouten identity (\ref{eq:schouten}), and derivatives of the field strength (\ref{eq:dfdf}) to prove the following identities\footnote{In deriving these we assumed Lorentzian signature. The single time-like direction then gives an extra minus sign when contracting Levi-Civita symbols.}:
\begin{equation}\begin{aligned}
\label{eq:finalsimplify}
	\widetilde{F}^{\mu\rho}\widetilde{F}^\nu_{~\rho} = -F^{\mu\rho}F^\nu_{~\rho}+\frac{1}{2}g^{\mu\nu}(F_{\rho\sigma}F^{\rho\sigma}) &= - 2R^{\mu\nu} + \frac{1}{4}g^{\mu\nu}(F_{\rho\sigma}F^{\rho\sigma})~, \\
	R_{\mu\nu\rho\sigma}(F^{\mu\nu}F^{\rho\sigma} - \widetilde{F}^{\mu\nu}\widetilde{F}^{\rho\sigma}) &= 8 R_{\mu\nu}R^{\mu\nu}~, \\
	(F_{\mu\nu}F^{\mu\nu})^2 - (F_{\mu\nu}\widetilde{F}^{\mu\nu})^2 &= 16 R_{\mu\nu}R^{\mu\nu}~.
\end{aligned}\end{equation}
These three relations are sufficient to rewrite all contractions involving the field strength in our $a_4(x)$ results purely in terms of the Riemann tensor.  This simplification is surprising because it would not work for generic four-derivative terms.  It was noted previously for the bosonic content of the gravity multiplet~\cite{Bhattacharyya:2012wz}.

From the argument above, we can write our $a_4(x)$ coefficients as
\begin{equation}
\label{eq:a4_ac}
	a_4(x) = \frac{c}{16\pi^2}W_{\mu\nu\rho\sigma}W^{\mu\nu\rho\sigma} - \frac{a}{16\pi^2} E_4~,
\end{equation}
for some constants $c$ and $a$, where the square of the Weyl tensor $W_{\mu\nu\rho\sigma}$ is
\begin{equation}
	W_{\mu\nu\rho\sigma}W^{\mu\nu\rho\sigma} = R_{\mu\nu\rho\sigma}R^{\mu\nu\rho\sigma} - 2 R_{\mu\nu}R^{\mu\nu} + \frac{1}{3}R^2~,
\end{equation}
and $E_4$ is the Euler density (also known as the Gauss-Bonnet term)
\begin{equation}
	E_4 = R_{\mu\nu\rho\sigma}R^{\mu\nu\rho\sigma} - 4 R_{\mu\nu}R^{\mu\nu} + R^2~.
\end{equation}
The $c,a$ constants can be identified as the central charges of the conformal anomaly in 4D.  In theories without dynamical gravity they are related to the renormalization group flow of the quantum field theory~\cite{Cardy:1988cwa,Komargodski:2011vj,Myers:2010tj,Elvang:2012st}.  

We now take the $a_4(x)$ results from~\S\ref{sec:heatkernel} and use the identities~(\ref{eq:finalsimplify}) to rewrite them in the form of~(\ref{eq:a4_ac}).  The results for the central charges of our theory (with a single graviton multiplet, $\mathcal{N}-2$ gravitino multiplets, $n_V$ vector multiplets and $n_H$ hyper multiplets) are listed in table~\ref{table:centralcharges}.

\bgroup
\def\arraystretch{1.5}
\begin{center}
\begin{table}[h]
\begin{tabular}{|c|c|c|}
\hline
\textbf{Fields} & $c$ & $a$ \\ \hline
Bosons & $\frac{1}{60}\left(137 + 12(\mathcal{N}-2) - 3 n_V + 2 n_H\right)$ & $\frac{1}{90}\left( 106 + 31(\mathcal{N}-2) + n_V + n_H\right)$ \\ \hline
Fermions & $-\frac{1}{60}\left(137 + 12(\mathcal{N}-2) - 3 n_V + 2 n_H\right)$ & $\frac{1}{360}\left( -589 + 41(\mathcal{N}-2) + 11 n_V - 19 n_H\right)$ \\ \hline
Total & $0$ & $\frac{1}{24}\left( -11 + 11(\mathcal{N}-2) + n_V - n_H\right)$ \\ \hline
\end{tabular}
\caption{Central charges $c$ and $a$ for the massless field content of a $\mathcal{N} \geq 2$ supergravity theory minimally coupled to the background gauge field.}
\label{table:centralcharges}
\end{table}
\end{center}
\egroup

As a check on these results we consider the special case of BPS black holes.  The near-horizon geometry for these spaces is AdS$_2\times S^2$, with non-zero components of the Riemann tensor
\begin{equation}
	R_{\alpha\beta\gamma\delta} = -\frac{1}{\ell^2}\left(g_{\alpha\gamma}g_{\beta\delta} - g_{\alpha\delta}g_{\beta\gamma} \right)~,\quad R_{ijkl} = \frac{1}{\ell^2}\left(g_{ik}g_{jl} - g_{il}g_{jk}\right)~,
\end{equation}
where $\ell$ is the radius of curvature of AdS$_2$ and $S^2$.  (The indices $\alpha,\beta,\gamma,\delta$ refer to AdS$_2$ and $i,j,k,l$ refer to $S^2$.)  It is straightforward to compute the curvature invariants
\begin{equation}
\label{eq:curveinv_ads2xs2}
	W_{\mu\nu\rho\sigma}W^{\mu\nu\rho\sigma} = 0~,\quad E_4 = -\frac{8}{\ell^4}~.
\end{equation}
If we combine these with the values of $c,a$ found in table~\ref{table:centralcharges} we reproduce the sum of the bulk and boundary contributions (for bosons and fermions separately) computed in~\cite{Keeler:2014bra,Larsen:2014bqa} exactly.

The results for $c,a$ in table~\ref{table:centralcharges} are fairly complicated when considering bosons and fermions separately.  However, the bosonic and fermionic values of $c$ for any full $\mathcal{N}=2$ multiplet exactly cancel, giving $c = 0$.  By simultaneously considering quadratic fluctuations of both the bosonic and fermionic fields in our theory, the $c$-anomaly vanishes for arbitrary $\mathcal{N} = 2$ multiplets.  The entire one-loop result depends only on the Euler density $E_4$:
\begin{equation}
	a_4(x) = -\frac{a}{16\pi^2}E_4~.
\end{equation}
This cancellation would not be noticed for supersymmetric black holes, since $W_{\mu\nu\rho\sigma}W^{\mu\nu\rho\sigma} = 0$ on AdS$_2\times S^2$ (\ref{eq:curveinv_ads2xs2}).

The cancellation of the $c$-anomaly is far from automatic.  For example, $c$ does not vanish in pure Einstein-Maxwell theory~\cite{Bhattacharyya:2012wz}, or equivalently for the bosonic part of our $\mathcal{N}=2$ supergravity multiplet.  $c$ and $a$ have been computed in many theories without dynamical gravity but rarely do these computations yield $c = 0$~\cite{Christensen:1978md,Anselmi:1997am,Anselmi:1997ys,Myers:2010xs}. For quantum field theories that can be described via the AdS/CFT correspondence the canonical situation is $c=a$ in the large $N$ limit~\cite{Balasubramanian:1999re,Henningson:1998gx,Freedman:1999gp,Gubser:1998vd,Benvenuti:2004dw}.
%%%%%%%%%%%%%%%%%%%%%%%%%%%%%%%%%%%%%%%%%%%%%%%%%%%%%%%%%%%%%%

\subsection{Black Hole Entropy}
\label{subsec:discussion_blackhole}

The logarithmic dependence of the black hole entropy is governed by
\begin{equation}
\label{eq:loggovern}
	\frac{\partial S}{\partial\log A_H} = \frac{1}{2}\left(C_\text{local} + C_\text{zm}\right)~,
\end{equation}
where $C_\text{local}$ is the constant term in the heat kernel $D(s)$ (\ref{eq:d(s)-k}) and $C_\text{zm}$ is an integer we add to account for zero modes.~\cite{Sen:2011ba,Banerjee:2011jp,Sen:2012cj,Sen:2012dw}.  

From the series representation of $K(x,x;s)$ in (\ref{eq:kseries}) it is clear that
\begin{equation}
\label{eq:clocal}
	C_\text{local} = \int d^4x\,\sqrt{-g}\,a_4(x) = - 2 a \chi~,
\end{equation}
where $\chi$ is the 4D Euler characteristic
\begin{equation}
	\chi = \frac{1}{32\pi^2}\int d^4x\,\sqrt{-g}\, E_4~.
\end{equation}
If we insert (\ref{eq:clocal}) into (\ref{eq:loggovern}) and employ the full value of the central charge $a$ from table~\ref{table:centralcharges}, we find the logarithmic correction to the black hole entropy
\begin{equation}
\label{eq:deltas}
	\delta S = \frac{\chi}{24}\bigg{(} 11-11(\mathcal{N}-2)-n_V+n_H\bigg{)}\log A_H + \frac{1}{2}C_\text{zm}\log A_H~.
\end{equation}
The logarithmic correction depends only on the Euler characteristic $\chi$ (as well as the zero mode correction $C_\text{zm}$).  This result is important because $\chi$ is a pure number that depends only on the topology of the black hole solution and not on any black hole parameters.

\subsection{Kerr-Newman Black Holes}
\label{subsec:discussion_kerrnewman}

The metric of a Kerr-Newman black hole parameterized by mass $M$, angular momentum $J$, and charge $Q$ is
\begin{equation}\begin{aligned}
	ds^2 &= -\frac{r^2+b^2\cos^2\psi-2Mr+Q^2}{r^2+b^2\cos^2\psi}dt^2 + \frac{r^2+b^2\cos^2\psi}{r^2+b^2-2Mr+Q^2}dr^2 + (r^2+b^2\cos^2\psi)d\psi^2 \\
	&\quad +\frac{(r^2+b^2\cos^2\psi)(r^2+b^2)+(2Mr-Q^2)b^2\sin^2\psi}{r^2+b^2\cos^2\psi}\sin^2\psi\, d\phi^2 \\
	&\quad + \frac{2(Q^2-2Mr)b}{r^2+b^2\cos^2\psi}\sin^2\psi \, dt\,d\phi~,
\end{aligned}\end{equation}
where $b = J/M$.  The horizon is located at
\begin{equation}
	r_H = M + \sqrt{M^2-Q^2-b^2}~,
\end{equation}
and the inverse temperature $\beta = \frac{1}{T}$ is
\begin{equation}
	\beta = \frac{2\pi M}{\sqrt{M^4-J^2-M^2Q^2}}\left(2M^2-Q^2+2\sqrt{M^4-J^2-M^2Q^2}\right)~.
\end{equation}
After Euclidean continuation $t\rightarrow -i\tau$ and interpreting $\tau$ as a periodic coordinate with period $\beta$ one can show that~\cite{Sen:2012dw}
\begin{equation}\begin{aligned}
\label{eq:intw2}
	\int d^4x\,\sqrt{-g}\,W_{\mu\nu\rho\sigma}W^{\mu\nu\rho\sigma} &= 64\pi^2 + \frac{\pi\beta Q^4}{b^5 r_H^4(b^2+r_H^2)}\bigg{[}4b^5r_H+2b^3r_H^3 \\
	&\quad +3(b^2-r_H^2)(b^2+r_H^2)^2\tan^{-1}\left(\frac{b}{r_H}\right)+3br_H^5\bigg{]}~.
\end{aligned}\end{equation}
This expression can be recast as a complicated function of two dimensionless ratios, e.g. $Q/M$ and $J/M^2$. In the extremal limit $M^2 = b^2 + Q^2$ the expression depends on only one of these ratios, but still in a very non-trivial way~\cite{Bhattacharyya:2012wz}.  In contrast, the integral of the Euler density is a pure number
\begin{equation}
\label{eq:inte4}
	\chi = \frac{1}{32\pi^2}\int d^4x\,\sqrt{-g}\,E_4 = 2~,
\end{equation}
for all values of the dimensionless ratios.

For generic field content the coefficient of the logarithmic correction to Kerr-Newman entropy resulting from (\ref{eq:a4_ac}) depends on the Weyl invariant and thus on all of the black hole parameters through the complicated expression in (\ref{eq:intw2}).  Our result in (\ref{eq:deltas}), however, demonstrates that when these black holes are interpreted as solutions to $\mathcal{N}\geq 2$ SUGRA there is dependence only on $\chi$ and thus the logarithmic corrections to Kerr-Newman entropy are universal: 
\begin{equation}
\label{eq:deltas_kn}
	\delta S = \frac{1}{12}\bigg{(} (11 + 6 C_\text{zm})-11(\mathcal{N}-2)-n_V+n_H\bigg{)}\log A_H~.
\end{equation}
The correction due to zero modes $C_\text{zm}$ depends on the setting.  Some important examples are:
\begin{itemize}
	\item \textbf{BPS black holes}: $C_\text{zm} = 2$~\cite{Sen:2011ba}. The background is spherically symmetric and preserves supersymmetry, giving rise to translational, rotational, and SUSY zero modes.
	
	\item \textbf{Extremal rotating Kerr-Newman}: $C_\text{zm} = -4$~\cite{Sen:2012cj}. The angular momentum breaks two of the rotational isometries and the background no longer preserves supersymmetry, leaving translational modes and one rotational mode.
	
	\item \textbf{Non-extremal rotating Kerr-Newman}: $C_\text{zm} = -1$~\cite{Sen:2012dw}.  The zero mode counting is the same as for the extremal case except with an additional correction due to the finite IR volume of integration.
	
\end{itemize}
For completeness we review the computation of $C_\text{zm}$ in Appendix~\ref{appendix_zeromodes}.

%%%%%%%%%%%%%%%%%%%%%%%%%%%%%%%%%%%%%%%%%%%%%%%%%%%%%%%%%%%%%%

\acknowledgments

We thank Ashoke Sen, Henriette Elvang, Pedro Lisb{\~a}o and Arash Ardehali for useful discussions. This work was supported in part by the U.S. Department of Energy under grant DE-FG02-95ER40899.

%%%%%%%%%%%%%%%%%%%%%%%%%%%%%%%%%%%%%%%%%%%%%%%%%%%%%%%%%%%%%%

\appendix
\section{Zero Modes}
\label{appendix_zeromodes}
We initially defined the heat kernel $D(s)$ in (\ref{eq:d(s)}) to include zero modes of $\Lambda$, so
\begin{equation}
\label{eq:d(s)_naive}
	D(s) = \sum_i e^{-s\lambda_i} = \sum_{\lambda_i \neq 0}e^{-s\lambda_i} + N_\text{zm}~,
\end{equation}
where $N_\text{zm}$ is the number of zero modes (i.e. the number of distinct eigenvalues of $\Lambda$ that are zero).  The contribution from the zero modes in (\ref{eq:d(s)_naive}) affects only the constant term $D_0$ and not any other terms in $D(s)$.

This contribution from zero modes must be reconsidered carefully. The schematic Euclidean path integral representation of the one-loop effective action (\ref{eq:pathint}) does not apply to zero modes, as the functional integral over the fields is no longer a Gaussian.  Instead, the zero mode piece of the path integral reduces to ordinary integrals over the symmetry groups that give rise to these zero modes.  These integrals depend on the scaling dimensions of the symmetry groups.  Contributions from zero modes were included in our local expressions but with an incorrect weight of 1, as in (\ref{eq:d(s)_naive}).  The correction due to the actual scaling dimension of the zero modes is
\begin{equation}
\label{eq:cscale}
	C_\text{zm} = -\sum_{i \in B} (\Delta_i - 1)N^{(i)}_\text{zm} + \sum_{i\in F} (2\Delta_i -1)N^{(i)}_\text{zm}~.
\end{equation}
where $\Delta$ is the scaling dimension of the field and $N_\text{zm}$ is the number of zero modes associated with that field~\cite{Sen:2011ba,Sen:2012cj,Keeler:2014bra}.  The fermionic zero modes have the opposite sign as bosonic zero modes to account for fermion spin statistics.  The fermionic scaling dimensions also count with double weight due to spin degeneracy.

The correct treatment of zero modes introduces the correction $C_\text{zm}$ in (\ref{eq:loggovern}).  As discussed in~\S 1 of~\cite{Sen:2012cj}, (\ref{eq:deltas_kn}) describes the logarithmic correction to the entropy in the microcanonical ensemble where $M$, $Q$, and $J$ are fixed. In general $C_\text{zm}$ can depend on how these quantities have been fixed.  This correction has been computed in many different cases~\cite{Sen:2012dw,Sen:2012cj,Sen:2011ba,Banerjee:2011jp}.  We collect these different results and present them compactly as
\begin{equation}
\label{eq:zm_final}
	C_\text{zm} = -(3+K) + 2 N_\text{SUSY} + 3 \delta~,
\end{equation}
where
\begin{equation}\begin{aligned}
	K &=
	\begin{cases}
		1 & \text{for $J_3$ fixed with $\vec{J}^2$ arbitrary} \\
		3 & \text{for $J_3 = \vec{J}^2 = 0$}
	\end{cases}~, \\
	N_\text{SUSY} &=
	\begin{cases}
		4 & \text{for BPS black holes} \\
		0 & \text{for non-BPS black holes}
	\end{cases}~, \\
	\delta &= 
	\begin{cases}
		1 & \text{for non-extremal black holes} \\
		0 & \text{for extremal black holes}
	\end{cases}~.
\end{aligned}\end{equation}

Scalars and spin-1/2 fermions have no zero modes.  Vector fields have scaling dimension $\Delta_1 = 1$, so there are no corrections due to vector zero modes.  All zero modes in the vector and gravitino multiplets are due to vector fields and thus these multiplets do not get corrected.  Therefore we only need consider the fields in the gravity multiplet.

The metric has scaling dimension $\Delta_2 = 2$ and $3+K$ zero modes.  There are 3 zero modes associated with translational invariance and $K$ zero modes associated with the number of rotational isometries of the black hole solution.

The fermionic zero modes have scaling dimension $\Delta_{3/2} = \frac{3}{2}$.  For BPS black hole solutions there are 4 SUSY zero modes, but there are no fermionic zero modes when the background does not preserve SUSY.

Non-extremal black holes have a finite temperature and thus we assume the inverse temperature $\beta$ scales with the length scale of the black hole, as opposed to the extremal limit where $\beta \rightarrow \infty$.  We thus have to consider a finite IR volume of integration, which gives a $3\delta$ contribution to (\ref{eq:zm_final}) that exactly cancels the translational zero modes for non-extremal black holes~\cite{Sen:2012dw}.

%%%%%%%%%%%%%%%%%%%%%%%%%%%%%%%%%%%%%%%%%%%%%%%%%%%%%%%%%%%%%%

\bibliographystyle{JHEP}
\bibliography{nonextremal_bibliography}

\end{document}